\documentclass[aps,pre,epsf,superscriptaddress,amsmath,amssymb,amsfonts,twocolumn,showpacs]{revtex4-1}

\usepackage{graphicx}
\usepackage{epsfig}
\usepackage{dcolumn}
\usepackage{bm}
\usepackage{braket}
\usepackage{amsmath}
\usepackage{mathtools}
\usepackage{color}
\usepackage{hyperref}
\newcommand{\abs}[1]{\left| #1 \right|} 

\usepackage{hyperref}
\usepackage{multirow}
\DeclareMathOperator{\sech}{sech} 

\begin{document}
\title{Correlated quantum dynamics of two quenched fermionic\\ impurities immersed in a Bose-Einstein Condensate}

\author{S. I. Mistakidis}
\affiliation{Center for Optical Quantum Technologies, Department of Physics, University of Hamburg, 
Luruper Chaussee 149, 22761 Hamburg Germany}
\author{L. Hilbig}
\affiliation{Center for Optical Quantum Technologies, Department of Physics, University of Hamburg, 
Luruper Chaussee 149, 22761 Hamburg Germany} 
\author{P. Schmelcher}
\affiliation{Center for Optical Quantum Technologies, Department of Physics, University of Hamburg, 
Luruper Chaussee 149, 22761 Hamburg Germany} \affiliation{The Hamburg Centre for Ultrafast Imaging,
Universit\"{a}t Hamburg, Luruper Chaussee 149, 22761 Hamburg, Germany}

\date{\today}

\begin{abstract} 

We unravel the nonequilibrium dynamics of two fermionic impurities immersed in a one-dimensional 
bosonic gas following an interspecies interaction quench. 
Monitoring the temporal evolution of the single-particle density of each species we reveal the existence of four distinct dynamical regimes. 
For weak interspecies repulsions both species either perform a breathing motion or the impurity density splits into two parts which interact 
and disperse within the bosonic cloud. 
Turning to strong interactions we observe the formation of dark-bright states within the mean-field approximation. 
However, the correlated dynamics shows that the fermionic density splits into two repelling density peaks which either travel towards the edges of the bosonic cloud 
where they equilibrate or they approach an almost steady state propagating robustly within the bosonic gas which forms density dips at the same location. 
For these strong interspecies interactions an energy transfer process from the impurities to their environment occurs at the many-body level, 
while a periodic energy exchange from the bright states (impurities) to the bosonic species is identified in the absence of correlations. 
Finally, inspecting the one-body coherence function for strong interactions enables us to conclude 
on the spatial localization of the quench-induced fermionic density humps.

\end{abstract}

\maketitle

\section{Introduction} 

Ultracold atoms offer a fertile testbed to monitor the nonequilibrium dynamics of 
quantum many-body (MB) systems due to their extraordinary degree of control. 
Recent experimental progress enables us, for instance, to adjust the interparticle interaction strength via Feshbach resonances \cite{Kohler,Chin}, 
and also realize multicomponent quantum gases consisting either of different isotopes \cite{Modugno} or different hyperfine states \cite{Myatt,Stenger} 
of the same species. 
Multicomponent quantum systems characterized by high population imbalanced components 
\cite{Wu,Heo,Cumby,Roati,Pilch,Schirotzek,Kohstall,Koschorreck,Zhang,Spethmann,Scazza,Robinson} 
have been a focal point of studies examining in particular the dressing of mobile impurities with the collective 
excitations of a surrounding MB system forming quasiparticles often referred to as polarons. 
As a consequence of this dressing mechanism a variety of the impurities properties, such as their effective mass and induced interactions can be 
strongly altered compared to the bare particle case. 
Beyond ultracold atoms applications include a multitude of different systems such as semiconductors \cite{Gershenson}, 
high temperature superconductors \cite{Dagotto}, doped Mott insulators \cite{doped_ins} and liquid Helium mixtures \cite{Bardeen,Baym}. 

The study of a mobile impurity immersed in an ultracold quantum gas has already provided numerous insights regarding Fermi 
\cite{Navon,Punk,Chevy,Cui,Pilati,Massignan1,Schmidt1,Schmidt2,Ngampruetikorn,Massignan2,Massignan3,Schmidt3,Burovski,Gamayun} and more recently 
Bose polarons \cite{Palzer,Tempere,Catani,Fukuhara,Scelle,Schmidtred,Ardila2,Grusdt,Artemis2,Guenther,Mayer,Ardila_new}. 
The majority of the theoretical investigations have focused on the stationary properties of these quasiparticle states and have been restricted to 
the mean-field (MF) approximation \cite{Astrakharchik,Cucchietti,Kalas,Bruderer1} and to the Fr{\"o}hlich model 
\cite{Sacha,Bruderer,Privitera,Casteels1,Casteels2,Kain}.   
Moreover, experimental evidences on the existence and dynamics of Fermi \cite{Scazza,Koschorreck,Kohstall} and Bose \cite{Jorgensen,Hu,Catani1,Fukuhara,Scelle,Yan_bose_polarons} 
polarons have triggered a new era of theoretical investigations in order to understand their nonequilibrium dynamics evincing also the necessity of taking into account 
higher-order correlations for an adequate description of the observed dynamics. 
Indeed, the impurities constitute a few-body system and correlation effects are expected to be well-pronounced. 
In this direction theoretical approaches that include correlations have been 
recently applied both to the Fermi~\cite{Mistakidis_fermi_pol} and mainly to the Bose polaron problem 
\cite{Li,Ardila1,Shchadilova,Rath,Grusdt1,Lemeshko,Kain1,Grusdt2,Mistakidis_bose_pol,Enss_beyond,Mistakidis_coh_state}, 
thereby enabling a first description of these quasiparticles also in the intermediate and strong interaction regime.

Despite the above-mentioned increasing amount of theoretical and experimental efforts, 
the nonequilibrium dynamics of such impurity systems is still largely unexplored. 
Especially in the case where more than a single impurity atom is involved and thus their interactions come into play. 
The simplest setup in order to advance our understanding of the emerging dynamics and reveal the correlation effects consists 
of two fermions immersed in a MB bosonic gas. 
In this case, the constituting particles obey different statistics \cite{Pethick_book,Lewenstein_book} and the impurities are non-interacting  
allowing us to avoid the additional complexity introduced by the intraspecies interactions. 
For this scenario, it would be particularly interesting to explore the dynamical response of the impurities for distinct interspecies interaction 
strengths and unveil the correlation properties of the quench-induced states. 
These might include the emergence of an orthogonality catastrophe for 
strong interactions \cite{Mistakidis_orthog_catastr}, the formation of dark-bright (DB) soliton complexes \cite{BF_brights,BF_brights1} or 
phase separation phenomena \cite{mistakidis_phase_sep,Erdmann_phase_sep}. 
To this end, we study the interspecies interaction quench dynamics of such a harmonically trapped Bose-Fermi (BF) mixture, both within a MF and a MB treatment, 
from weak to strong repulsions. 
To track the correlated quantum dynamics of the BF mixture we employ the Multi-Layer Multi-Configuration Time-Dependent Hartree 
Method for Atomic Mixtures (ML-MCTDHX) \cite{MLX}, which is a non-perturbative variational method capturing all interparticle correlations. 

Regarding the ground state properties of the system we show that a phase separation process between the two species occurs, both at the MF and the MB level, for 
interspecies interaction strengths larger than the bosonic intraspecies ones \cite{phase_sep_BF,phase_sep_BF_th,phase_sep_BF_th1,phase_sep_BF_th2}. 
To induce the dynamics we apply an interspecies interaction quench starting from a weakly repulsive ground state.  
Depending on the postquench interspecies coupling we realize four different dynamical regimes. 
For weak postquench interactions both species remain miscible and perform a breathing motion \cite{breathing_BF,phase_sep_BF_th2}. 
Increasing the postquench interaction strength the single-particle density of the impurities splits into two effectively repelling density peaks that are 
seen to disperse within the bosonic gas. 
This behavior is more pronounced at the MB level. 
In the strong interspecies interaction regime and focusing on the MF approximation we observe the spontaneous generation of two DB solitary waves, 
with the bright solitons emerging in the fermionic species and the dark ones appearing in the bosonic gas. 
These structures interact repelling and approaching one another throughout evolution. 
In sharp contrast, in the MB description these solitary waves soon after their formation are pushed towards the edges of the bosonic gas remaining there 
in the course of the evolution, exhibiting also a gradually decaying amplitude. 
Entering the very strong interspecies interaction regime we again observe the formation of DB structures in both approaches. 
In contrast to their MF evolution, the DB solitary waves formed at the MB level tend, for later evolution times, to approach an almost steady state. 
Within the latter two strongly interacting regimes it is found that an energy exchange process, from the impurities to the bosonic bath, takes place in the presence of correlations. 
Moreover, monitoring the one-body coherence function reveals the appearance of Mott-like correlations \cite{Sherson,darkbright_beyond,mistakidis_phase_sep,Erdmann_phase_sep} 
between the emergent bright solitary waves indicating their tendency for localization. 
Finally inspecting the dynamics of the fermionic two-body reduced density matrix we unveil that depending on the postquench 
interaction strength the two fermionic impurities either behave almost independently or experience a weak attraction for very strong interspecies couplings. 

The present work is structured as follows. 
Section \ref{sec:theory} presents our setup and the different observables that are used for the identification of the correlated character of the observed dynamics. 
The nonequilibrium dynamics induced by an interspecies interaction quench of the particle imbalanced BF mixture is discussed in Sec. \ref{sec:quench}.  
We summarize and provide an outlook in Section \ref{sec:conclusions}. 
Finally, in Appendix \ref{sec:convergence} we provide further details of our numerical simulations 
and demonstrate their convergence exemplarily.

\section{Theoretical Framework}\label{sec:theory}

\subsection{Setup}\label{sec:hamiltonian}

We consider a particle imbalanced BF mixture consisting of $N_F=2$ spin polarized fermionic impurities and $N_B=100$ bosons 
which constitute the majority species. 
The mixture is further assumed to be mass balanced, namely $M_B=M_F\equiv M$, while each species is trapped in the same external harmonic oscillator 
potential of frequency $\omega_B=\omega_F=\omega$. 
Such an approximately mass balanced mixture can be experimentally realized by considering e.g. a mixture of isotopes of $^{7}$Li and $^{6}$Li \cite{Kempen} 
or $^{171}$Yb and $^{172}$Yb \cite{Honda,Takasu}.       
The resulting MB Hamiltonian of the system reads 
\begin{equation}
\begin{split}
	\label{eq:hamiltonian}
	H = \sum_{\sigma = F,B} \sum_{i = 1}^{N_{\sigma}}\left[-\frac{\hbar^2}{2M}\left(\frac{\mathrm{d}}{\mathrm{d}x_i^{\sigma}}\right)^2 + 
	\frac{1}{2} M \omega (x_i^{\sigma})^2\right] \\+ g_{BB} \sum_{i<j} \delta(x_i^B - x_j^B) + g_{BF} \sum_{i = 1}^{N_{F}} \sum_{j = 1}^{N_{B}} \delta(x_i^F-x_j^B).
\end{split}
\end{equation}
We operate within the ultracold regime and therefore $s$-wave scattering is the dominant interaction process. 
Consequently both the inter- and intraspecies interactions are described by contact interactions whose effective one-dimensional 
coupling strength \cite{Olshanii} is ${g_{\sigma \sigma'}} =\frac{{2{\hbar ^2}{a^s_{\sigma \sigma'}}}}{{\mu a_ \bot ^2}}{\left( {1 - {\left|{\zeta (1/2)} \right|{a^s_{\sigma \sigma'}}}/{{\sqrt 2 {a_ \bot }}}} \right)^{ -
1}}$. 
Here, $\sigma,\sigma'=B, F$ for bosons or fermions respectively and $\mu=\frac{M}{2}$ is the corresponding reduced mass. 
The transversal length scale reads ${a_\bot } = \sqrt{\hbar /{\mu{\omega _ \bot }}}$ with ${{\omega _ \bot }}$ denoting the transversal confinement frequency while 
${a^s_{\sigma \sigma'}}$ is the three-dimensional $s$-wave scattering length within ($\sigma=\sigma'$) or between ($\sigma \neq \sigma'$) the two distinct species. 
Moreover, $s$-wave scattering is forbidden for spin-polarized fermions \cite{Pethick_book,Lewenstein_book}, due to the antisymmetry 
of the fermionic wavefunction, and thus intraspecies interactions within the fermionic species are neglected. 
We remark that $g_{\sigma\sigma'}$ can be experimentally tuned either via ${a^s_{\sigma \sigma'}}$ by means of Feshbach resonances \cite{Kohler,Chin} 
or by adjusting ${{\omega _ \bot }}$ via confinement-induced resonances \cite{Olshanii}. 

The Hamiltonian of Eq. (\ref{eq:hamiltonian}) is rescaled in units of $\hbar  \omega_{\perp}$. 
As a result the corresponding length, time, and interaction strength are expressed in terms of
$\sqrt{\frac{\hbar}{M \omega_{\perp}}}$, $\omega_{\perp}^{-1}$ and 
$\sqrt{\frac{\hbar^3 \omega_{\perp}}{M}}$ respectively. 

In the present work, we prepare our system in the ground state of the Hamiltonian (\ref{eq:hamiltonian}) within the weak intra- and interspecies interaction 
regime, namely $g_{BB}=0.5$ and $g_{BF}=0.1$. 
Therefore the two species are miscible, i.e. their one-body densities spatially overlap. 
Recall that in the absence of a trap species separation takes place when $g^2_{BF}>g_{BB}$, otherwise the two species 
overlap \cite{phase_sep_BB,phase_sep_BF,phase_sep_BF_th}.   
Another important remark here is that for sufficiently strong trapping, a scenario not considered herein, 
the above condition needs to be modified \cite{phase_sep_BB} namely $g_{BF}$ should become substantially larger than $g_{BB}$ in order to overcome 
the implicitly miscibility favoring effect of the trap. 
To trigger the nonequilibrium dynamics of the BF mixture we suddenly change at $t=0$ the interspecies interaction strength towards a larger value of repulsion, 
e.g. $g_{BF}=1.5$, which favors species immiscibility and let the system evolve in time. 
Below, we first briefly discuss the ground state properties of the system [Sec.  \ref{sec:ground_state}] and then analyze in detail the quench-induced 
dynamics [Sec.  \ref{sec:quench}] of the two fermionic impurities immersed in the bosonic gas.

\subsection{Wavefunction ansatz}

To calculate the stationary properties of the BF mixture and most importantly the quench-induced nonequilibrium dynamics we solve 
the corresponding MB Schr{\"o}dinger equation utilizing ML-MCTDHX \cite{MLX}.   
This method is based on expanding the MB wavefunction with respect to a time-dependent and variationally optimized basis which allows us 
to take into account both the inter- and the intraspecies correlations of the BF mixture. 
Accordingly, in order to incorporate interspecies correlations we expand the MB wavefunction in terms of $D$ different species 
functions, $\Psi^{\sigma}_k (\vec x^{\sigma};t)$, for each component. 
In this notation the spatial $\sigma=B,F$-species coordinates are $\vec x^{\sigma}=\left( x^{\sigma}_1, \dots, x^{\sigma}_{N_{\sigma}} \right)$ 
and the number of the $\sigma$-species atoms is $N_{\sigma}$. 
As a result, the MB wavefunction $\Psi_{MB}$ acquires the form of a truncated Schmidt decomposition \cite{Horodecki} of rank $D$ 
\begin{equation}
\Psi_{MB}(\vec x^F,\vec x^B;t) = \sum_{k=1}^D \sqrt{ \lambda_k(t) }~ \Psi^F_k (\vec x^F;t) \Psi^B_k (\vec x^B;t).    
\label{Eq:WF}
\end{equation}
In the following, we shall also refer to the Schmidt coefficients $\lambda_k(t)$ as the natural species 
populations of the $k$-th species function. 
Moreover, the system is assumed to be entangled \cite{Roncaglia} or interspecies correlated if at least two distinct $\lambda_k(t)$ 
are nonzero since in this latter case $\Psi_{MB}$ is not a direct product of two states. 

Furthermore in order to include intraspecies correlations into our MB ansatz we expand each of 
the species functions $\Psi^{\sigma}_k (\vec x^{\sigma};t)$ in terms of the determinants and permanents of $d_{\sigma}$ distinct 
time-dependent fermionic and bosonic single-particle functions (SPFs) $\varphi_1^{\sigma},\dots,\varphi_{d_{\sigma}}^{\sigma}$ respectively. 
Consequently, $\Psi^{\sigma}_k (\vec x^{\sigma};t)$ is expressed as follows
\begin{equation}
\begin{split}
&\Psi_k^{\sigma}(\vec x^{\sigma};t) = \sum_{\substack{l_1,\dots,l_{d_{\sigma}} \\ \sum l_i=N}} c_{k,(l_1,
\dots,l_{d_{\sigma}})}(t)\sum_{i=1}^{N_{\sigma}!} {\rm sign}(\mathcal{P}_i) ^{\zeta}\times \\ & \mathcal{P}_i
 \left[ \prod_{j=1}^{l_1} \varphi_1^{\sigma}(x_j;t) \cdots \prod_{j=1}^{l_{d_{\sigma}}} \varphi_{d_{\sigma}}^{\sigma}(x_{K(d_{\sigma})+j};t) \right].  
 \label{Eq:SPF}
 \end{split}
\end{equation} 
The index $\zeta=0,1$ stands for the case of bosons and fermions respectively. 
$\rm{sign}(\mathcal{P}_i)$ refers to the sign of the corresponding permutation with $\mathcal{P}$ denoting the permutation 
operator which exchanges the particle positions $x_{\nu}^{\sigma}$, $\nu=1,\dots,N_{\sigma}$ within the SPFs. 
The symbol $K(r)\equiv \sum_{\nu=1}^{r-1}l_{\nu}$, where $l_{\nu}$ is the occupation of the $\nu$th SPF and $r\in\{1,2,\dots,d_{\sigma}\}$. 
Also, $c_{k,(l_1,\dots,l_{d_{\sigma}})}(t)$ are the time-dependent expansion coefficients of a certain determinant for fermions or permanent for bosons. 
Moreover, the eigenfunctions of the $\sigma$-species one-body reduced density matrix 
$\rho_\sigma^{(1)}(x,x^\prime;t)=\langle\Psi_{MB}(t)|\hat{\Psi}^{\sigma \dagger}(x)\hat{\Psi}^\sigma(x^\prime)|\Psi_{MB}(t)\rangle$ 
are the so-called natural orbitals $\phi^{\sigma}_i(x;t)$.  
Here $\hat{\Psi}^{F}(x)$ and $\hat{\Psi}^{B}(x)$ refer to the fermionic and bosonic field operators respectively.  
It is worth mentioning at this point that the natural orbitals are related with the SPFs via a unitary transformation that diagonalizes 
$\rho_\sigma^{(1)}(x,x^\prime;t)$ when it is expressed in the basis of SPFs, for more details see also \cite{MLX,MLB}. 
The resulting eigenvalues of $\rho_\sigma^{(1)}(x,x^\prime;t)$ are termed natural populations $n^{\sigma}_i(t)$. 
In the following we will refer to the bosonic or fermionic subsystem as intraspecies correlated if more than one (for bosons) or $N_F$ (for fermions) 
eigenvalues are macroscopically occupied. 
Otherwise the corresponding subsystem will be termed fully coherent or Hartree-Fock correlated respectively.  

To obtain the ML-MCTDHX equations of motion \cite{MLX} we follow e.g. the Dirac-Frenkel variational principle \cite{Frenkel,Dirac} for the 
MB ansatz given by Eqs.~(\ref{Eq:WF}), (\ref{Eq:SPF})]. 
This procedure results in $D^2$ linear differential equations of motion for the coefficients $\lambda_i(t)$ being coupled to a set of 
$D[$ ${N_B+d_B-1}\choose{d_B-1}$+${d_F}\choose{N_F}$] non-linear integro-differential equations for the species functions and $d_F+d_B$ 
integro-differential equations for the SPFs. 
Another important feature of ML-MCTDHX is that it enables us to operate within different approximation 
orders. 
For instance, we can retrieve the corresponding MF equation \cite{Pethick_book,Lewenstein_book} 
of the BF mixture in the limit of $D=d_B=1$ and $d_F=N_F$. 
In this latter case the MB wavefunction ansatz boils down to the following MF product state  
\begin{equation}
\begin{split}
&\Psi_{MF}(\vec x^{B},\vec x^{F};t) =
	\frac{1}{\sqrt{N_F!}}\prod_{j=1}^{N_B}\varphi_1^B(x_j^{B};t)\times \\& \sum_{i=1}^{N_{F}!} {\rm sign}(\mathcal{P}_i) \mathcal{P}_i
\left[\varphi_1^F(x_1^{F};t) \cdots \varphi_{N_{F}}^F(x_{N_{F}}^{F};t) \right].  
\label{Eq:MF}
\end{split}
\end{equation} 
Utilizing a variational principle, such as the Dirac-Frenkel variational principle \cite{Frenkel,Dirac}, we obtain the corresponding 
system of coupled ($N_F+1$) MF equations of motion \cite{Pethick_book,BF_brights1} that govern the dynamics of the BF mixture 
\begin{equation}
\begin{split}
 &i\frac{\partial \phi_1^B(x;t)}{\partial t}= \bigg[-\frac{1}{2m} \frac{\partial^2}{\partial x^2}+\frac{1}{2} M\omega x^2
 +g_{BB}N_B \\& \times\abs{\phi_1^B(x;t)}^2+g_{BF}\sum_{i=1}^{N_F} \abs{\phi_i^F(x;t)}^2 \bigg] \phi_1^B(x;t), \\  
 &i\frac{\partial \phi_j^F(x;t)}{\partial t}= \bigg[-\frac{1}{2m} \frac{\partial^2}{\partial x^2}+\frac{1}{2} M\omega x^2
 \\&+g_{BF}N_B\sum_{i=1}^{N_F} \abs{\phi_1^B(x;t)}^2 \bigg] \phi_j^F(x;t). 
 \end{split}
\end{equation}
Here the first equation describes the dynamics of the bosonic gas and corresponds to the well-known Gross-Pitaevskii equation of motion for a BF mixture.  
Moreover the equation for $\phi_j^F(x;t)$ characterizes the time-evolution of the $j$-th fermionic orbital and represents the set 
of $j=1,\dots,N_F$ Hartree-Fock equations of motion. 
Within the MF approximation only the trivial Hartree-Fock intraspecies correlations are taken into account stemming from the existence of $N_F$ distinct SPFs. 
A next interesting reduction of the method is the so-called species mean-field (SMF) approximation \cite{darkbright_beyond,Mistakidis_bose_pol}. 
Here, the entanglement between the species is ignored but the correlations within each of the species are included. 
Then, the total wavefunction of the system acquires the following tensor product form 
\begin{equation} 
    |\Psi(t)\rangle_{SMF}= |\Psi^{B}_1(t)\rangle \otimes |\Psi^{F}_1(t)\rangle. 
    \label{eq:SMF}
\end{equation}
Indeed, the system's wavefunction is described by only one species wavefunction, i.e. $\ket{\Psi_k^{B}(t)}=\ket{\Psi_{k}^{F}(t)}=0$ for $k\neq 1$ which 
is expanded in terms of the time-dependent basis of Eq. (\ref{Eq:SPF}) consisting of different time-dependent variationally 
optimized SPFs.

\subsection{Correlation measures}\label{observables}

To investigate the role of intraspecies correlations, at the one-body level, during the interaction quench dynamics of the BF mixture we utilize the 
spatial first order coherence function \cite{Naraschewski,density_matrix,mistakidis_phase_sep}
\begin{align}
g^{(1)}_\sigma(x,x^\prime;t)=\frac{\rho_\sigma^{(1)}(x,x^\prime;t)}{\sqrt{\rho_\sigma^{(1)}(x;t)\rho_\sigma^{(1)}(x^\prime;t)}}.\label{one_body_cor}
\end{align} 
In this expression, $\rho_\sigma^{(1)}(x,x^\prime;t)=\langle\Psi(t)|\hat{\Psi}^{\sigma \dagger}(x)\hat{\Psi}^\sigma(x^\prime)|\Psi(t)\rangle$ is 
the $\sigma$ species one-body reduced density matrix while $\rho_\sigma^{(1)}(x;t)\equiv\rho_\sigma^{(1)}(x,x^\prime=x;t)$ denotes the corresponding 
one-body density.  
Also, $\hat{\Psi}^{B}(x)$ and $\hat{\Psi}^{F}(x)$ refer to the corresponding bosonic and fermionic field operators at position $x$ satisfying the standard 
commutation and anti-commutation relations respectively \cite{Pethick_book}. 
Most importantly, the one-body coherence function $|g^{(1)}_{\sigma}(x,x';t)|$ takes values in the interval $[0,1]$ and provides a degree of the deviation of 
the MB state from a product state for the set of coordinates $x$, $x'$. 
Indeed, two distinct spatial regions denoted e.g. by $R$ and $R'$, where $R \cap R' = \varnothing$, 
exhibiting $|g^{(1)}_{\sigma}(x,x';t)|= 1$, when $x\in R$ and $x'\in R'$, are termed fully coherent. 
In this case, the absence of one-body correlations in these regions can be inferred. 
However, when $|g^{(1)}_{\sigma}(x,x';t)|<1$ for $x\in R$ and $x'\in R'$ the regions are referred to as partially incoherent. 
Here, the aforementioned inequality signifies the emergence of one-body intraspecies correlations. 
Another interesting situation is when full coherence occurs within a spatial region $R$, i.e. when
$|g^{(1),\sigma}(x,x';t)|^2 \approx 1$ $x,x'\in R$, 
while perfect incoherence takes place between different spatial regions $R$, $R'$, when $|g^{(1),\sigma}(x,x';t)|^2\approx 0$ with 
$x\in R$, $x'\in R'$ and $R \cap R' = \varnothing$. 
This latter case suggests the emergence of Mott-like correlations \cite{Sherson,darkbright_beyond,mistakidis_phase_sep} into 
the system and often indicates the spatial localization of the underlying structures building upon the respective one-body density \cite{mistakidis_phase_sep,Erdmann_phase_sep,darkbright_beyond}.  

To monitor the effective interactions between the non-interacting fermionic impurities in the course of the evolution we 
employ their relative distance \cite{Mistakidis_fermi_pol} 
\begin{equation}
\label{eq:distance}
\begin{split}
D(t)=\frac{\int dx_1 dx_2 |x_1-x_2| \rho^{(2)}_{FF}(x_1,x_2;t)}{
\braket{\Psi_{MB}(t)|\hat{N}_{F} \left(\hat{N}_{F} -1\right)|\Psi_{MB}(t)}},    
\end{split}
\end{equation} 
Here, $\rho^{(2)}_{FF}(x_1,x_2;t)=\bra{\Psi_{MB}(t)}\Psi^{F \dagger}(x_1)\Psi^{F \dagger}(x_2)\Psi^{F}(x_1)\\\Psi^{F}(x_2)\ket{\Psi_{MB}(t)}$ 
is the diagonal two-body intraspecies reduced density matrix. 
This quantity provides the probability of finding the two fermions at the positions $x_1$ and $x_2$ at time $t$ \cite{mistakidis_phase_sep,Erdmann_phase_sep}. 
Also, $\hat{N}_{F}$ is the number operator that measures the number of fermions. 
Most importantly, $D(t)$ can be directly probed experimentally by performing {\it in-situ} spin-resolved single-shot measurements on 
the fermionic state \cite{Jochim2}. 
In particular, each image provides an estimate of $D(t)$ between the fermionic impurities given that their position uncertainty is relatively small \cite{Jochim2}. 
Then, $D(t)$ is obtained by averaging over several such images. 

To quantify the degree of both intra- and interspecies correlations during the nonequilibrium dynamics we calculate the fragmentation 
of the $\sigma$-species and the entanglement between the species of the BF mixture \cite{Koutentakis_prob_fer,Erdmann_tunel,Erdmann_phase_sep}. 
This investigation allows us to infer about the proximity of the MB state [Eq.~(\ref{Eq:WF})] to a MF one [see Eq.~(\ref{Eq:MF})]. 
The presence of interspecies correlations or entanglement is designated by the values of the higher than the first Schmidt 
coefficients, i.e. $\lambda_k(t)$ with $k>1$. 
Recall that $\lambda_k(t)$ are the eigenvalues of the species reduced density matrix
$\rho^{N_{\sigma}} (\vec{x}^{\sigma}, \vec{x}'^{\sigma};t)=\int d^{N_{\sigma'}} x^{\sigma'} \Psi^*_{MB}(\vec{x}^{\sigma}, 
\vec{x}^{\sigma'};t) \Psi_{MB}(\vec{x}'^{\sigma},\vec{x}^{\sigma'};t)$, with $\vec{x}^{\sigma}=(x^{\sigma}_1, \cdots, 
x^{\sigma}_{N_{\sigma-1}})$, and $\sigma\neq \sigma'$ [see also Eq. (\ref{Eq:WF})]. 
Accordingly, the system is termed species entangled or interspecies correlated when more than a single eigenvalues of $\rho^{N_{\sigma}}$ 
are macroscopically populated, otherwise it is non-entangled [see also the discussion around Eq. \ref{Eq:WF})]. 
A commonly used measure to identify the degree of species entanglement constitutes the Von-Neumann 
entropy \cite{Erdmann_phase_sep,Erdmann_tunel,Catani} 
\begin{align}
S_{VN}(t)=-\sum\limits_{k=1}^D \lambda_{k}(t)\ln[\lambda_k(t)].\label{eq:entropy} 
\end{align} 
Note here that within the MF limit $S_{VN}(t)=0$ holds since $\lambda_1(t)=1$, while for a MB state where more than one $\lambda_k$ 
contribute $S_{VN}(t)\neq0$. 

To unveil the fragmented or intraspecies correlated nature of each species we resort to the eigenvalues, $n^{\sigma}_i(t)= \int d x~ \left| \phi^{\sigma}_i(x;t) \right|^2$, 
of the $\sigma$-species one-body reduced density matrix, $\rho^{(1)}_{\sigma}(x,x';t)$ \cite{density_matrix,MLX}. 
Note that $\phi_i^{\sigma}((x;t))$ are the so-called natural orbitals. 
It can be shown that when $\Psi_{MB}(\vec x^F,\vec x^B;t) \to \Psi_{MF}(\vec x^F,\vec x^B;t)$, 
see also Eq.~(\ref{Eq:WF}) and Eq.~(\ref{Eq:MF}), 
the fermionic and bosonic natural populations obey $\sum_{i=1}^{N_{F}}n_i^{F}(t)=1$, $n_{i>N_{F}}^{F}(t)=0$ and $n_1^{B}(t)=1$, $n_{i>1}^{B}(t)=0$ respectively.  
As a result, when more than $N_{F}$ (one) fermionic (bosonic) natural orbitals are significantly populated the system is referred to as fragmented 
and the corresponding degree of fragmentation can be quantified as follows  
\begin{equation}
F_{F}(t) = 1 - \sum\limits_{i=1}^{N_{F}}n_i^F(t)~~~ {\rm and}~~~F_{B}(t) =1 - n_1^B(t). \label{fragmentation}
\end{equation} 
These constitute theoretical tools for the identification of the occupation of the $d_{F} - N_{F}$ and $d_B>1$ least occupied fermionic and bosonic  
natural orbitals respectively, and thus of the deviation of the MB state from the MF one when $F_{F}(t)>0$ and $F_{B}(t)>0$.  

\begin{figure}[ht]
 	\includegraphics[width=0.5\textwidth]{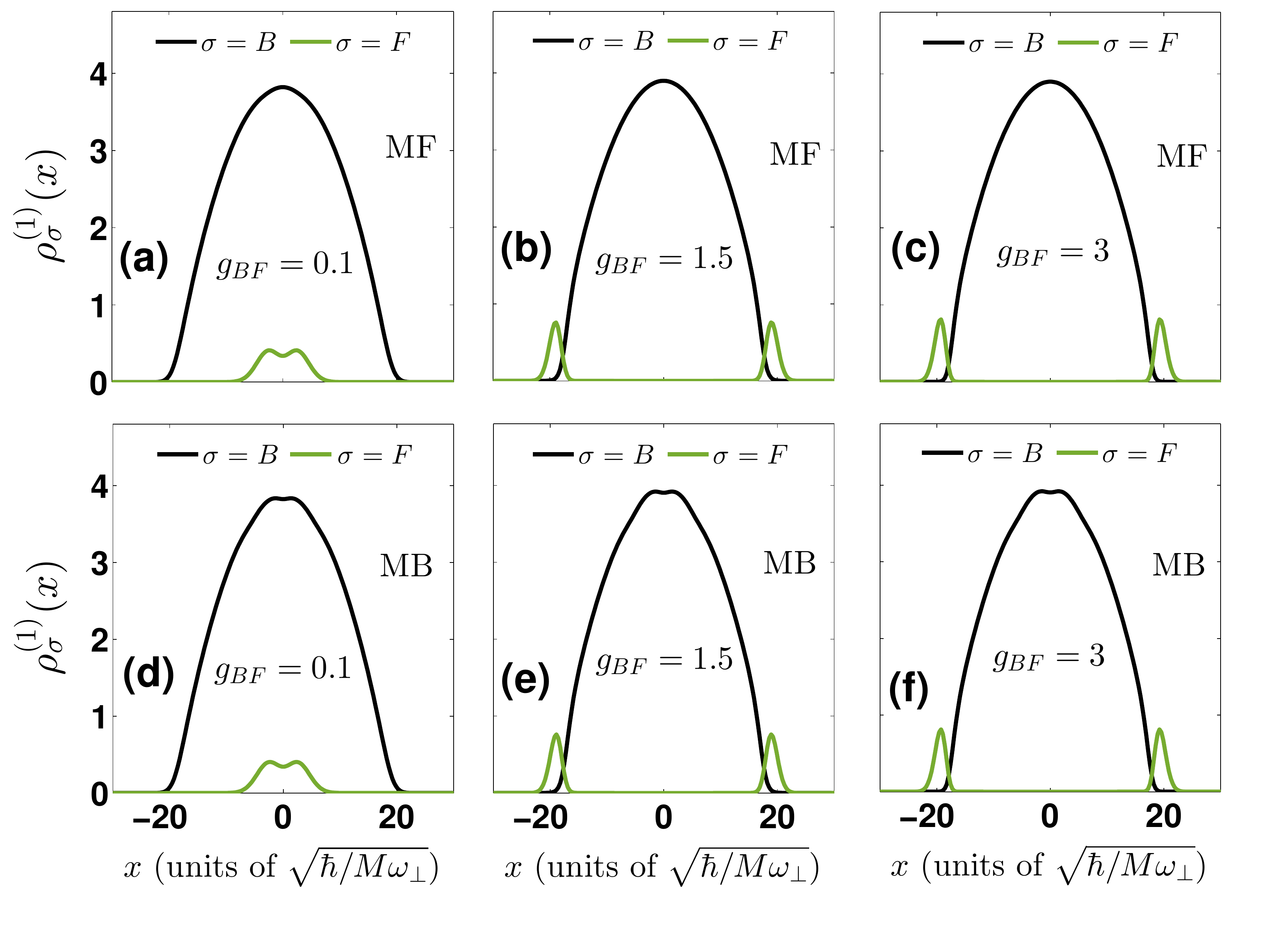}
 	\caption{One-body density $\rho^{(1)}_\sigma(x)$ of the ground state of the $\sigma$-species of the BF mixture for varying 
 	interspecies repulsions $g_{BF}$ (see legend) within (a), (b), (c) the MF approximation and (d), (e), (f) the MB approach. 
 	The BF mixture consists of $N_B=100$ bosons and $N_F=2$ fermions with $g_{BB}=0.5$ and it is trapped in a harmonic oscillator potential with $\omega=0.1$. 
 	Densities are expressed in units of $\sqrt{M\omega_{\perp}/\hbar}$.}
 	\label{fig:ground} 
 \end{figure}

\section{Quench Induced Dynamics}\label{sec:quench}

In the following we investigate the interspecies interaction quench dynamics of the particle imbalanced BF mixture. 
The emergent nonequilibrium dynamics is explored both in the MF approximation [see Eq. (\ref{Eq:MF})] and in the MB 
approach [see Eq.~(\ref{Eq:WF}) and Eq.~(\ref{Eq:SPF})]. 
Throughout this work we consider a BF mixture consisting of $N_B=100$ bosons and $N_F=2$ spin polarized fermionic impurities. 
In particular, the system is initialized in its weakly interacting ground state characterized by $g_{BB}=0.5$ and $g_{BF}=0.1$, 
unless it is stated otherwise. 
To induce the dynamics we perform an interspecies interaction quench from $g_{BF}=0.1$ to a strongly interacting state such that $g_{BF}>g_{BB}$ and 
thus a phase separation process between the two components is favored (see also the discussion below). 

\subsection{Ground state of the BF mixture}\label{sec:ground_state}

Before delving into the nonequilibrium dynamics of the BF mixture it is instructive to briefly analyze its ground state properties 
for fixed intraspecies interactions, $g_{BB}=0.5$, and varying interspecies interactions $g_{BF}$. 
The mixture is confined in a harmonic oscillator potential of frequency $\omega=0.1$ and it is prepared in its corresponding repulsively interacting ground state 
as described by the Hamiltonian of Eq.~(\ref{eq:hamiltonian}). 
To obtain the ground state of Eq. (\ref{eq:hamiltonian}) we use either imaginary time propagation or improved relaxation \cite{MLX,MLB} within ML-MCTDHX. 
\begin{figure}[ht]
 	\includegraphics[width=0.5\textwidth]{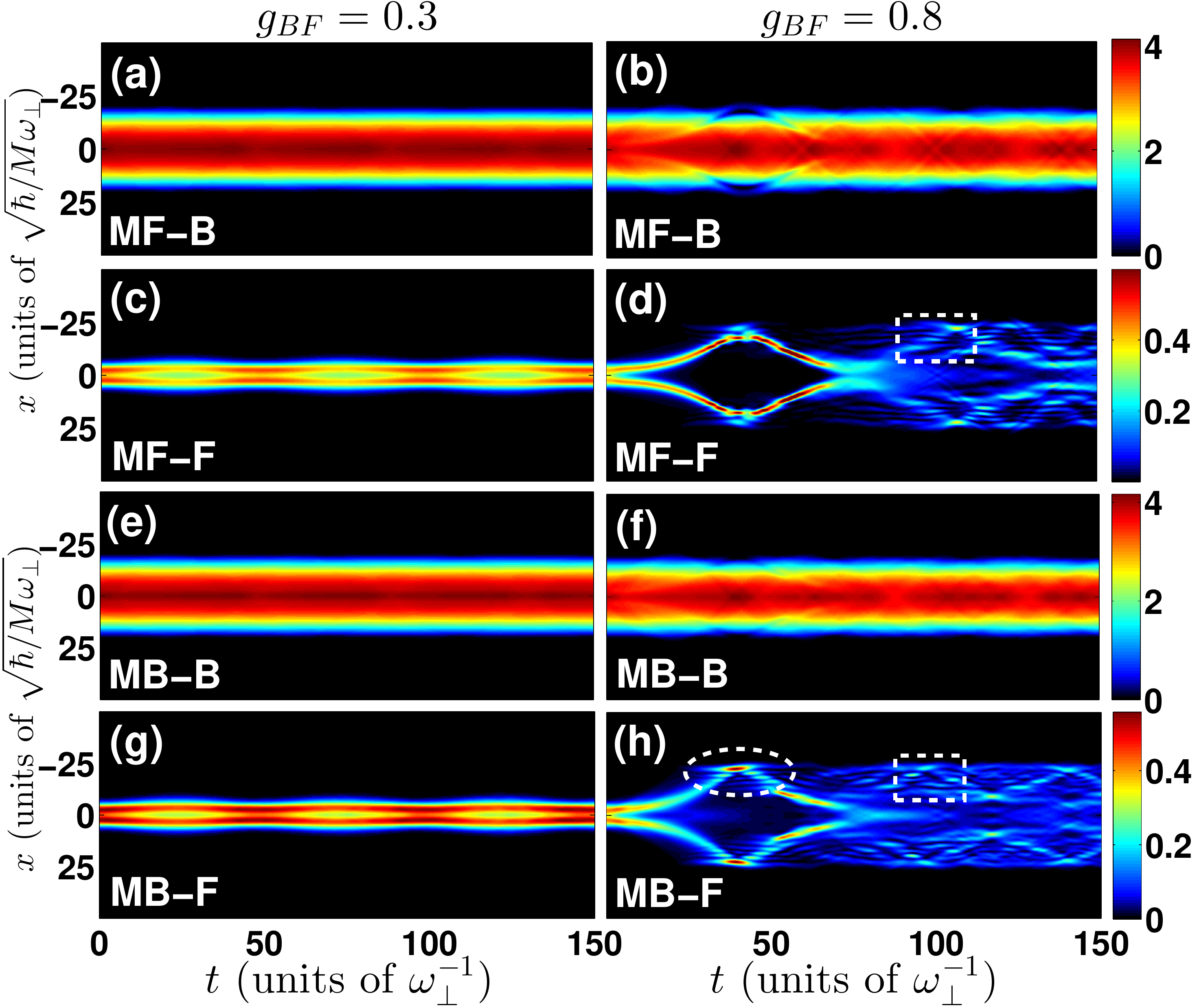}
 	\caption{Time-evolution of the $\sigma=B,F$-species (see legends) one-body density, $\rho_{\sigma}^{(1)}(x;t)$, of the BF mixture 
 	following an interspecies interaction quench within (a)-(d) the MF approximation and (e)-(h) the MB approach. 
 	The postquench interspecies interaction strengths correspond to (a), (c), (e), (g) $g_{BF}=0.3$ and (b), (d), (f), (h) 
 	$g_{BF}=0.8$.
 	Dashed rectangles in (d), (h) mark the presence of several local minima in $\rho_{F}^{(1)}(x;t)$, while the dashed ellipse in (h) 
 	indicates the splitting of $\rho_{F}^{(1)}(x;t)$ for $x<0$.   
 	The system consists of $N_B=100$ bosons and $N_F=2$ fermions trapped in a harmonic oscillator potential and it is initialized in 
 	its ground state for $g_{BB}=0.5$ and $g_{BF}=0.1$.}
 	\label{fig:density} 
 \end{figure} 

 \begin{figure}[ht]
 	\includegraphics[width=0.45\textwidth]{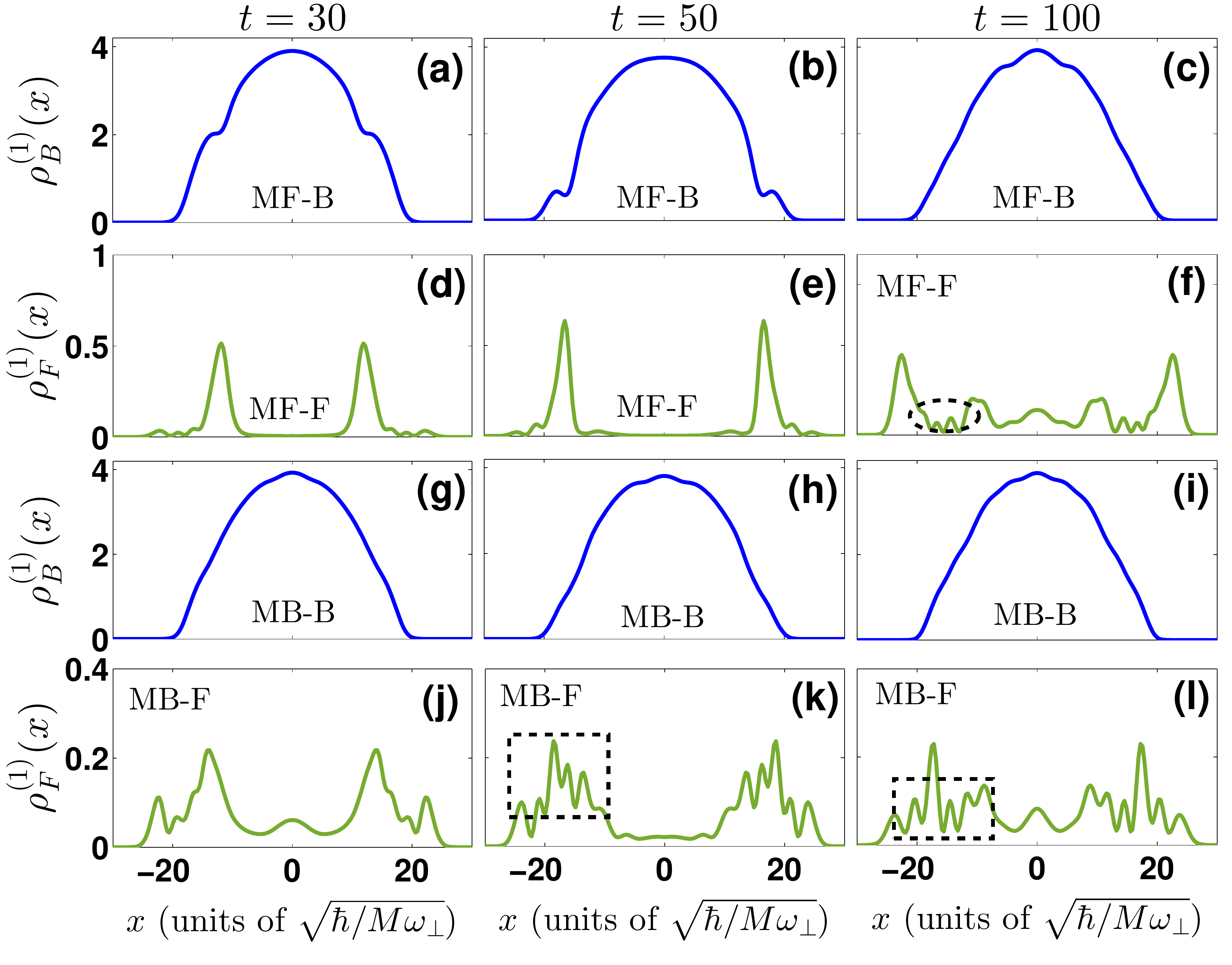}
 	\caption{One-body density profiles, $\rho^{(1)}_{\sigma}(x;t)$, of the $\sigma=B,F$ species of the BF mixture 
 	at different time instants following an interspecies interaction quench from $g_{BF}=0.1$ to $g_{BF}=0.8$ (see legends).   
 	(a)-(c) [(d)-(f)] present $\rho^{(1)}_{B}(x;t)$ [$\rho^{(1)}_{F}(x;t)$] within the MF approximation. 
 	(g)-(i) [(j)-(l)] show $\rho^{(1)}_{B}(x;t)$ [$\rho^{(1)}_{F}(x;t)$] determined in the MB approach. 
 	The dashed ellipse in (f) and the dashed rectangles in (k), (l) indicate the presence of local minima in $\rho^{(1)}_{F}(x;t)$ for 
 	$x<0$ determined at the MF and the MB level respectively. 
 	The remaining system parameters are the same as in Fig. \ref{fig:density}. 
 	The densities are given in units of $\sqrt{M\omega_{\perp}/\hbar}$.}
 	\label{fig:profiles_first} 
\end{figure} 

To explore the ground state properties of the BF mixture we employ the $\sigma$-species 
single-particle density $\rho_\sigma^{(1)}(x)$ [see also Eq. (\ref{one_body_cor})]. 
Figure \ref{fig:ground} presents $\rho_\sigma^{(1)}(x)$ within the MF approximation [Figs. \ref{fig:ground} (a)-(c)] 
and on the MB level [see Figs. \ref{fig:ground} (d)-(g)] for different interspecies interaction strengths and fixed $g_{BB}=0.5$. 
In all cases $\rho_B^{(1)}(x)$ possesses the form of a Thomas-Fermi profile and it is almost 
insensitive to the value of $g_{BF}$. 
In particular, for an increasing interspecies repulsion the Thomas-Fermi profile becomes slightly more compressed and accordingly its maximum at $x=0$ acquires 
a larger value [hardly discernible in Fig. \ref{fig:ground}]. 
Note also that due to the particle imbalance the bosons, being the majority species, exhibit a broader single-particle density distribution 
when compared to the fermions.  
Thus, $\rho_F^{(1)}(x)$ shows a much smaller amplitude than $\rho_B^{(1)}(x)$ and its shape depends crucially on $g_{BF}$ \cite{phase_sep_BF,phase_sep_BF_th,phase_sep_BF_th1,phase_sep_BF_th2}. 
More specifically, within the weakly interspecies interaction regime, $g_{BF}=0.1$, $\rho_F^{(1)}(x)$ resides well inside the edges of $\rho_B^{(1)}(x)$ and therefore the 
two species are miscible since their spatial overlap is finite [Figs.~\ref{fig:ground}(a), (d)]. 
For increasing interspecies repulsion, e.g. $g_{BF}=1.5$, $\rho_F^{(1)}(x)$ splits into two density branches each one located either at the right or the left edge 
of $\rho_B^{(1)}(x)$ respectively [Figs.~\ref{fig:ground}(b), (e)]. 
This behavior indicates the immiscible character of the mixture for strong $g_{BF}$, 
i.e. the spatial overlap between $\rho_B^{(1)}(x)$ and 
$\rho_F^{(1)}(x)$ is almost zero, a character that remains as such for even stronger $g_{BF}$ [see Figs.~\ref{fig:ground}(c), (f)]. 
Also, by carefully inspecting $\rho_F^{(1)}(x)$ for these strong interactions we can deduce that for an increasing $g_{BF}$ the density peaks of $\rho_F^{(1)}(x)$ 
become slightly more localized and their relative distance slightly decreases [hardly discernible in Figs.~\ref{fig:ground}(b) and (c)]. 
Another important observation here is that both $\rho_B^{(1)}(x)$ and $\rho_F^{(1)}(x)$ are essentially the same within the MF approximation and the MB approach. 
Indeed the inclusion of correlations causes only small deviations in the corresponding ground state density profiles. 
For instance, a shallow local minimum appears in $\rho_B^{(1)}(x)$ around $x=0$ within the MB approach which is absent at the MF level, e.g. compare 
Figs.~\ref{fig:ground}(b) and (e). 
The existence of such a local minimum in $\rho_B^{(1)}(x)$ suggests a minor involvement of higher-excited states of the harmonic oscillator for the correct characterization of the ground state.

\subsection{Single-particle density evolution}\label{subsec:density}

\begin{figure}[ht]
 	\includegraphics[width=0.46\textwidth]{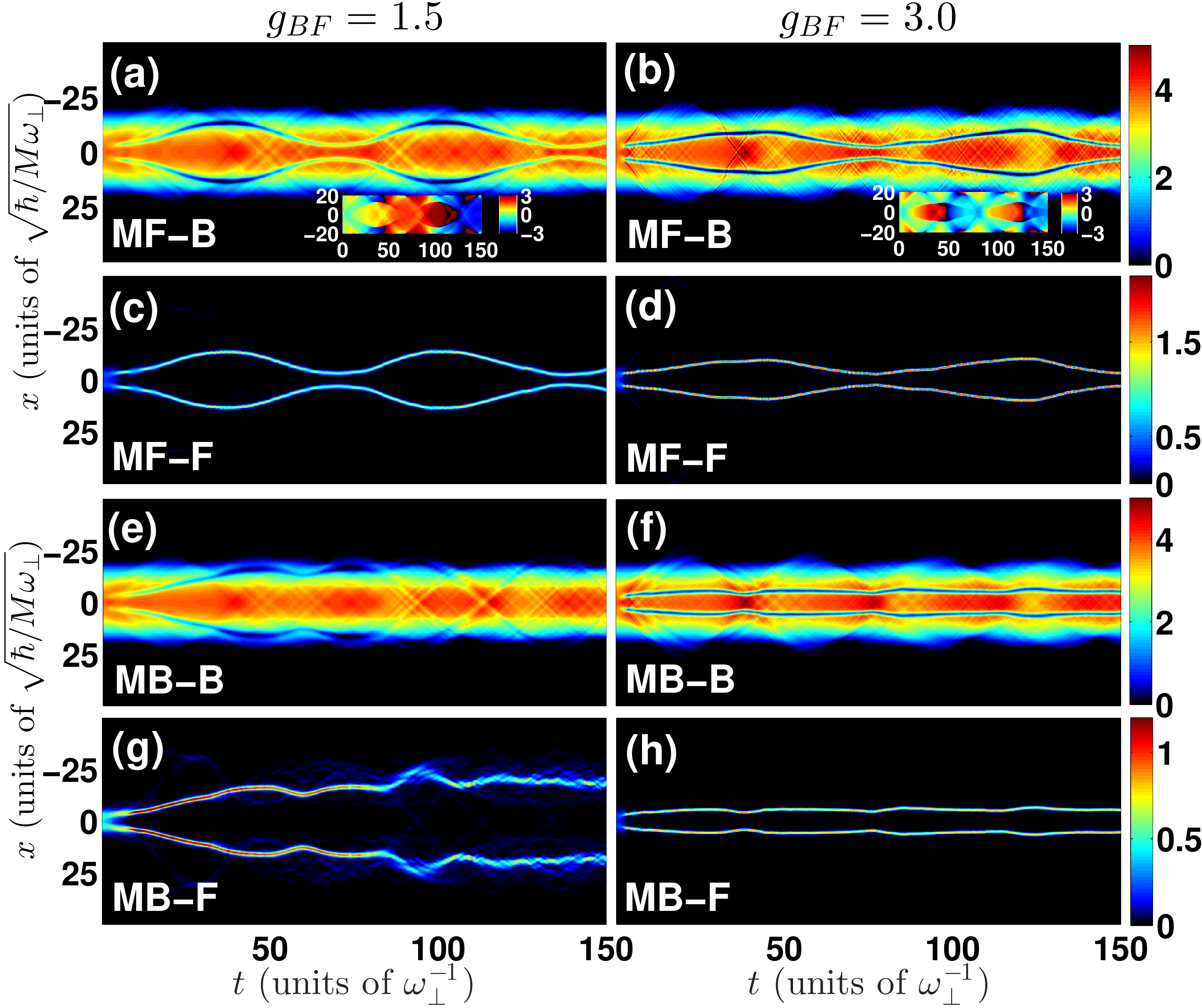}
 	\caption{Spatiotemporal evolution of the $\sigma=B,F$-species one-body density, $\rho_{\sigma}^{(1)}(x;t)$, 
 	of the BF mixture upon considering an 
 	interspecies interaction quench within (a)-(d) the MF approximation and (e)-(h) the MB approach (see legends). 
 	The postquench interspecies interaction strength corresponds to (a), (c), (e), (g) $g_{BF}=1.5$ and (b), (d), (f), (h) $g_{BF}=3$.  
 	The insets in (a), (b) show the corresponding phases during the dynamics.    
 	The system consists of $N_B=100$ bosons and $N_F=2$ fermions trapped in a harmonic oscillator potential and it is initialized in 
 	its ground state characterized by $g_{BB}=0.5$ and $g_{BF}=0.1$.}
 	\label{fig:density_strong} 
 \end{figure}

Having analyzed the ground state properties of the BF mixture for increasing interspecies interactions, we next investigate its nonequilibrium dynamics 
following a sudden change of $g_{BF}$ from $g_{BF}=0.1$ towards stronger repulsive interspecies interactions. 
Figures \ref{fig:density} and \ref{fig:density_strong} show $\rho^{(1)}_{\sigma}(x;t)$ of the bosonic and fermionic species both within the MF and the MB level for different 
characteristic postquench interspecies interaction strengths. 
As it can be seen, the dynamics of the mixture on the single-particle level can be categorized into four distinct interaction regions. 
Most importantly, the structures building upon $\rho^{(1)}_{\sigma}(x;t)$ differ considerably between the MF and the MB approach, especially when $g_{BF}>g_{BB}$. 
In particular, for a quench within the weak interspecies interaction regime such that $g_{BF}<g_{BB}$, e.g. $g_{BF}=0.3$, both $\rho^{(1)}_{B}(x;t)$ and 
$\rho^{(1)}_{F}(x;t)$ exhibit a breathing motion [see Figs.~\ref{fig:density}(a), (c)] 
characterized by a frequency $\omega_{br}^B\approx 0.21\equiv 2\omega$ 
and $\omega_{br}^F\approx 0.16$ respectively \cite{breathing_BF,phase_sep_BF_th2}. 
Note that the total external potential of the fermionic impurities is, to a very good approximation, the effective potential created by the 
harmonic oscillator and the density of bosons \cite{Mistakidis_orthog_catastr,effective_pot_MF}, 
i.e. $V_{eff}=\frac{1}{2} m \omega^2 x^2+ g_{BF} \rho_B^{(1)}(x)$, 
with $\rho_B^{(1)}(x)$ being the bosonic single-particle density at $t=0$. 
Hence, assuming the Thomas-Fermi approximation for $\rho^{(1)}_B$ we obtain the following effective trapping frequency of the impurities $\omega_{eff}=\omega\sqrt{1-\frac{g_{BF}}{g_{BB}}}$ and 
therefore their corresponding effective breathing frequency would be $\omega_{br}^{eff,F}=2\omega_{eff}=0.16$ which is indeed in a 
very good agreement with the numerically obtained $\omega_{br}^F$. 
We remark that this effective potential approximation is adequate only for small interspecies interactions where the entanglement between the species is weak 
[see Sec.~\ref{subsec:degree_cor}] and the impurities do not probe the edges of the bosonic cloud, 
see also~\cite{Mistakidis_bose_pol,Mistakidis_orthog_catastr} for more details. 
Moreover, $\rho^{(1)}_{F}(x;t)$ resides within $\rho^{(1)}_{B}(x;t)$ throughout the evolution indicating the miscible character of the dynamics.   
This behavior of $\rho^{(1)}_{\sigma}(x;t)$ occurs both at the MF [Figs.~\ref{fig:density}(a), (c)] and at the MB level 
[Figs.~\ref{fig:density}(e), (g)] since the degree of intra- and 
interspecies correlations is negligible here [see also Sec.~\ref{subsec:degree_cor}]. 

Increasing the postquench interaction strength $g_{BF}$ to a value larger than $g_{BB}$ 
the impurities undergo a much more involved dynamics compared to the weakly interacting case, while the majority species performs a breathing motion [Figs.~\ref{fig:density}(b), (f)] 
again with $\omega_{br}^B\approx 0.21$. 
Focusing on the MF approximation we observe that directly after the quench $\rho^{(1)}_{F}(x;t)$ splits into two density 
branches [Figs.~\ref{fig:profiles_first}(d), (e)] that acquire finite momenta and travel towards the edges of the bosonic 
cloud [Figs.~\ref{fig:density}(b), (d) and Figs.~\ref{fig:profiles_first}(a), (b)].  
Note that the appearance of these density branches is caused by the interaction quench which imports energy into the system. 
Reaching the edges of $\rho^{(1)}_{B}(x;t)$ these density humps of $\rho^{(1)}_F(x;t)$ are reflected back towards the trap center. 
Subsequently, around $t\approx 70$, they collide at $x=0$ and then show a dispersive behavior. 
In particular, directly after the collision $\rho^{(1)}_{F}(x;t)$ splits into several localized density humps, 
reflecting in this way its spatial delocalization, that are seen to propagate predominantly within   
$\rho^{(1)}_{B}(x;t)$ during evolution, see Fig.~\ref{fig:density}(d). 
This dispersive character of $\rho^{(1)}_{F}(x;t)$ is clearly captured and shown in the corresponding profile snapshots, 
see for instance the dashed ellipse in Fig.~\ref{fig:profiles_first}(f) and the relevant dashed rectangle in Fig.~\ref{fig:density}(d).  
In turn these localized humps indicate that $\rho^{(1)}_{F}(x;t)$ is in a superposition of several lower-lying excited states of the external potential into which the impurities are trapped. 
It is also important to note here that since the impurities probe the edges of the Thomas-Fermi radius of $\rho^{(1)}_{B}(x;t)$,
the effective potential picture previously introduced is not sufficient to describe the observed dynamics. 
The same overall phenomenology is also observed at the MB level.
However, important differences between the two approaches can be noticed especially in the time-evolution of $\rho^{(1)}_{F}(x;t)$. 
The most important feature here is that each of the two initially formed density branches [Fig.~\ref{fig:profiles_first}(j)] split into
several density peaks [Figs.~\ref{fig:profiles_first}(k), (l)] of significantly lower intensity. 
This further splitting, that is in contrast to the MF outcome, occurs when each of the fermionic density humps approaches the edges of the BEC medium 
and persists even upon their return towards the trap center [see here the dashed ellipse in Fig.~\ref{fig:density}(h) and also the 
dashed rectangle in Fig.~\ref{fig:profiles_first}(k)]. 
This behavior, that is absent within the MF approximation, is attributed to correlations present in the interaction of the impurities with the bosonic bath. 
Moreover, the dispersive character of $\rho^{(1)}_{F}(x;t)$ is more pronounced at the MB level with the number of localized
density humps being larger as compared to the MF description of the dynamics  
[compare e.g. Figs.~\ref{fig:profiles_first}(f) and (l)]. 
The latter observation indicates that the impurities are in a superposition of energetically higher excited states 
as compared to the MF scenario. 
Note also that $\rho^{(1)}_{F} (x;t)$ shown e.g. in Fig.~\ref{fig:density}(h) results from the correlated MB approach and 
a corresponding interpretation in terms of $V_{eff}$ in this case provides only a crude picture of the impurity dynamics 
since $V_{eff}$ does not contain any information about correlations. 

\begin{figure}[ht]
 	\includegraphics[width=0.46\textwidth]{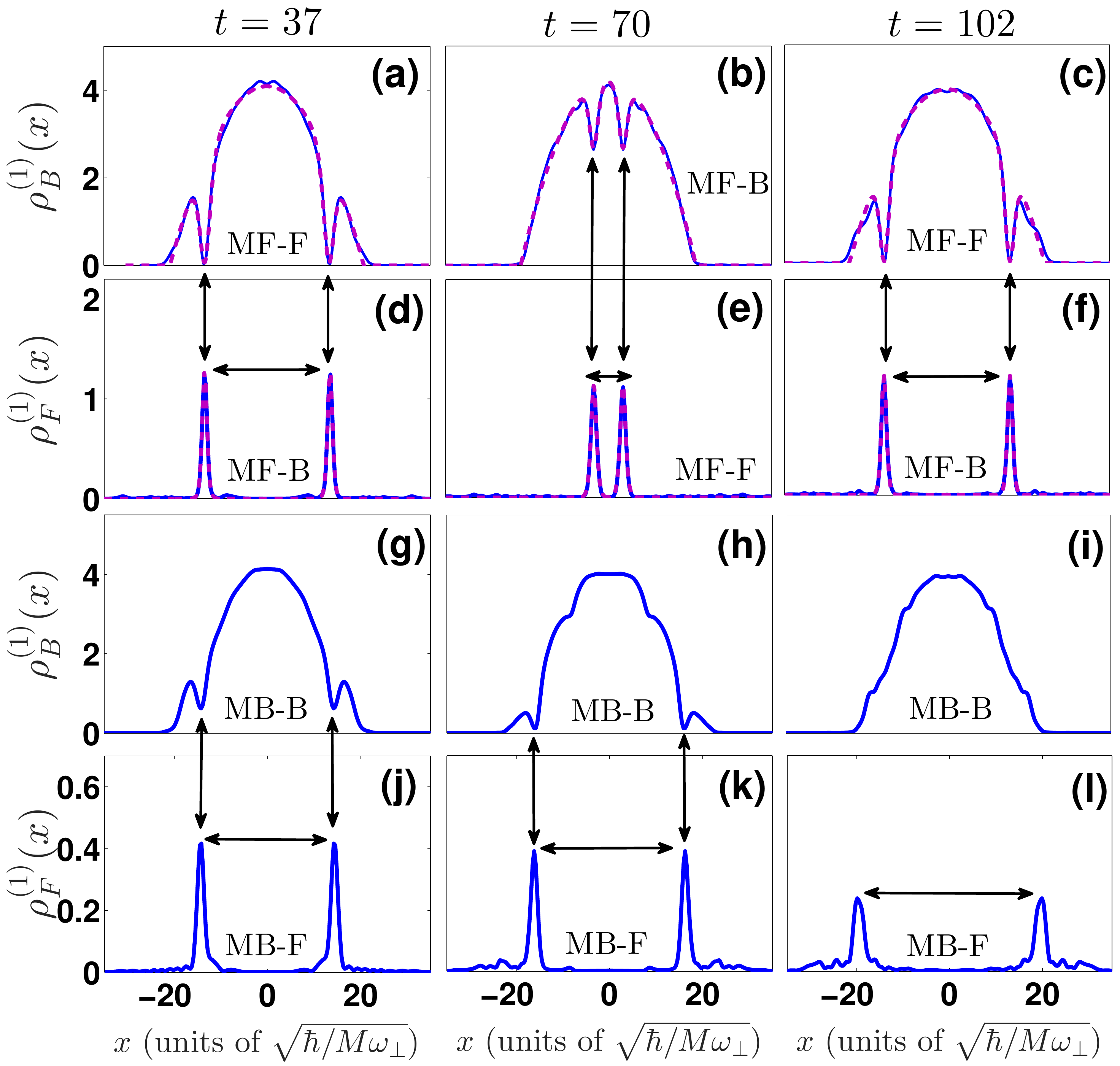}
 	\caption{Snapshots of the $\sigma=B,F$ species (see legend) one-body density, $\rho^{(1)}_{\sigma}((x;t))$, of the BF mixture 
 	at selected time instants (see legends) after an interspecies interaction quench from $g_{BF}=0.1$ to $g_{BF}=1.5$.   
 	(a)-(c) [(d)-(f)] show $\rho^{(1)}_{B}((x;t))$ [$\rho^{(1)}_{F}((x;t))$] within the MF approximation. 
 	(g)-(i) [(j)-(l)] illustrate $\rho^{(1)}_{B}((x;t))$ [$\rho^{(1)}_{F}((x;t))$] within the MB approach. 
 	The vertical double arrows indicate the corresponding DB pairs while the horizontal arrows show the relative distance between the 
 	fermions. 
 	The remaining system parameters are the same as in Fig. \ref{fig:density_strong}. 
 	The densities are expressed in units of $\sqrt{M\omega_{\perp}/\hbar}$.}
 	\label{fig:profiles} 
\end{figure}

Turning to even stronger postquench interaction strengths, e.g. $g_{BF}=1.5$, the dynamical behavior of both the fermionic and the 
bosonic species changes drastically compared to the above discussed cases. 
Most importantly, the dynamics is severely different between the MF the MB approach as can be deduced 
by comparing Figs.~\ref{fig:density_strong}(a), (c) to Figs.~\ref{fig:density_strong}(e), (g) respectively. 
Inspecting $\rho^{(1)}_{B}(x;t)$ [Fig.~\ref{fig:density_strong}(a)] and $\rho^{(1)}_{F}(x;t)$ [Fig.~\ref{fig:density_strong}(c)] within the MF approximation we observe the 
spontaneous generation of two localized density peaks in $\rho^{(1)}_{F}(x;t)$ robustly propagating within the BEC medium repelling and attracting 
one another, thus performing an oscillatory motion.   
These density humps are accompanied by the simultaneous formation of two density dips in $\rho^{(1)}_{B}(x;t)$ located at the same spatial regions and filled by the density peaks 
of the fermionic species. 
Such a filling mechanism resembles the formation of DB solitons in defocusing media which in our case oscillate within the parabolic trap with a 
period $T_{DB}\approx 69$ [see Fig.~\ref{fig:density_strong}(c)]. 
Let us note in passing that for the specific chosen parameters of our system the spontaneous generation of DB solitons occurs at postquench interspecies 
interaction strengths larger than $g_{BF}=0.9$. 
To provide evidence that indeed the structures building upon $\rho^{(1)}_{B}(x;t)$ possess a dark soliton character the spatio-temporal 
evolution of the relevant phase is illustrated in the inset of Fig.~\ref{fig:density_strong}(a). 
Evidently, the entities formed exhibit a phase jump being multiple of $\pi$. 
To further support our above-mentioned arguments regarding the DB character of the quench generated structures 
we present in Figs.~\ref{fig:profiles}(a)-(c) and (d)-(f) 
profile snapshots of the single-particle densities $\rho^{(1)}_{B}(x;t)$ and $\rho^{(1)}_{F}(x;t)$ respectively at $t=37$, $t=70$ and $t=102$ for a quench to $g_{BF}=1.5$. 
To identify the occurrence of DB states we employ the exact in the so-called integrable limit 
single DB soliton solution~\cite{Yan_DB,DB_oscil,Katsimiga_DB1,Katsimiga_DB2}.  
In this case the corresponding wavefunction ansatz for a dark and a bright soliton state reads    
$\Psi_B^l((x,t))=\cos\phi \tanh \left[D \left(x-x_0(t)\right)\right]+i \sin \phi$ and 
$\Psi_F^l((x,t))=B \sech \left[D \left(x-x_0(t)\right) \right] e^{ikx +i\theta(t)}$. 
Here, $\cos\phi$ and $B$ denote the amplitude of the dark and the bright soliton respectively and $D$ is their common inverse width. 
Moreover, $\sin\phi$ is the dark soliton's velocity, $x_0(t)$ is the soliton's center, $k=D\tan\phi$ is the constant wavenumber of the bright 
soliton, $\theta(t)$ its phase and $l$ indexes the number of the DB pair. 
Since in our case two DB soliton pairs are spontaneously generated, we further assume that the wavefunction 
ansatz that describes such a two DB state is approximately given by $\Psi_B((x;t))= \Psi_B^1((x,t)) \Psi_B^2((x,t))$ 
for the dark solitons developed in the bosonic component and $\Psi_F(x,t)=\Psi_F^1(x,t)+\Psi_F^2(x,t)$ for the bright states formed in the fermionic one. 
We use these expressions, and in particular $\abs{\Psi_B(x,t)}^2$ and $\abs{\Psi_F(x,t)}^2$, for the fits in the density profiles shown with dashed lines in Figs.~\ref{fig:profiles}(a)-(f). 
Evidently, a remarkably good agreement between the MF numerical simulations and the fitted profiles occurs 
[see also Table \ref{table}], thus 
verifying that indeed the quench-induced structures possess a DB solitary wave character. 

Having identified the formation of DB states at the MF level next let us inspect 
how the nonequilibrium dynamics is altered in the presence of 
correlations.  
As it can be seen both $\rho^{(1)}_{B}(x;t)$ [Figs.~\ref{fig:density_strong}(e)] and $\rho^{(1)}_{F}(x;t)$ 
[Figs.~\ref{fig:density_strong}(g)] are significantly different compared to their MF counterparts. 
In particular, after the quench $\rho^{(1)}_{F}(x;t)$ breaks into two distinct localized density peaks 
which travel towards the edges of $\rho^{(1)}_{B}(x;t)$. 
Within this time interval, $0<t<40$, the density peaks of $\rho^{(1)}_{F}(x;t)$ are accompanied 
by density dips in $\rho^{(1)}_{B}(x;t)$ resembling this way DB states. 
This situation is more evident in the corresponding profile snapshots of the densities [Figs.~\ref{fig:profiles}(g), (h), (j), (k)] where the density 
dips building upon the bosonic bath are filled by the density peaks formed in the fermionic species. 
However, when these localized pairs reach the edges of $\rho^{(1)}_{B}(x;t)$, instead of returning towards the trap center, 
they remain at the edges of the cloud while oscillating locally.
As can be deduced by focusing our attention to the fermionic species [Fig.~\ref{fig:density_strong}(g)] 
the observed oscillations are rather irregular and result in localized states whose amplitude is almost half 
the initial one [see also Fig.~\ref{fig:profiles}(l)]. Notice also that the initially formed density dips in $\rho^{(1)}_{B}(x;t)$
are hardly visible for evolution times $t>100$, a result that is clearly seen in the profile snapshot presented in  
Fig.~\ref{fig:profiles}(i). The above observations suggest that the DB character of the states formed 
is significantly altered by the presence of correlations. 
This alteration is captured by the decaying amplitude of the dark states formed in the BEC. 
	
\begin{table}
\begin{tabular}{|p{0.3cm}|p{1.1cm}||p{1.48cm}|p{1.48cm}|p{1.58cm}||p{1.58cm}|}
 \hline
 \multicolumn{4}{|c|}{{\bf Two DB soliton pair characteristics}} \\
 \hline \hline 
  
 &~~~~&~~~$t=37$ &~~~~~~$t=102$\\
 \hline \hline
 \parbox[t]{2mm}{\multirow{4}{*}{\rotatebox[origin=c]{90}{Dark~~~}}}
 &~~$\cos\phi$  &~~~~$1.341$ &~~~~~~~~~~~~$1.320$\\
 &~~~~$D$ &~~~~$0.669$ &~~~~~~~~~~~~$0.749$\\
 &~~$\sin \phi$  &~~~~$0.160$ &~~~~~~~~~~~~$0.122$\\
 &~~~~$x_{0}$  &~~~~$13.390$ &~~~~~~~~~~~~$13.580$\\
 &~~~~$R_{TF}$  &~~~~$20.790$ &~~~~~~~~~~~~$21.060$\\
 \hline 
 \hline
 \parbox[ht]{2mm}{\multirow{4}{*}{\rotatebox[origin=c]{90}{~~Bright}}}
 &~~~~$B$  &~~~~$0.878$ &~~~~~~~~~~~~$0.909$ \\
 &~~~~$D$ &~~~~$0.669$ &~~~~~~~~~~~~$0.749$\\
 &~~~~$x_0$  &~~~~$13.450$ &~~~~~~~~~~~~$13.570$\\
 \hline
\end{tabular}
\caption{Two DB soliton pair characteristics referring to different time instants of the dynamics of both species 
shown in Figs. \ref{fig:density_strong} (c) and Figs. \ref{fig:profiles} (a), (d), (c) and (f).  
The parameters $\cos \phi$ and $B$ refer to the amplitude of the dark and bright solitons respectively generated in the $\sigma=B$ and $\sigma=F$ species. 
Additionally, $R_{TF}$ denotes the Thomas-Fermi radius of the BEC background. 
Note also that in all cases the accuracy of the fitting is $0.95$.}
\label{table}
\end{table}

\begin{figure}[ht]
 	\includegraphics[width=0.48\textwidth]{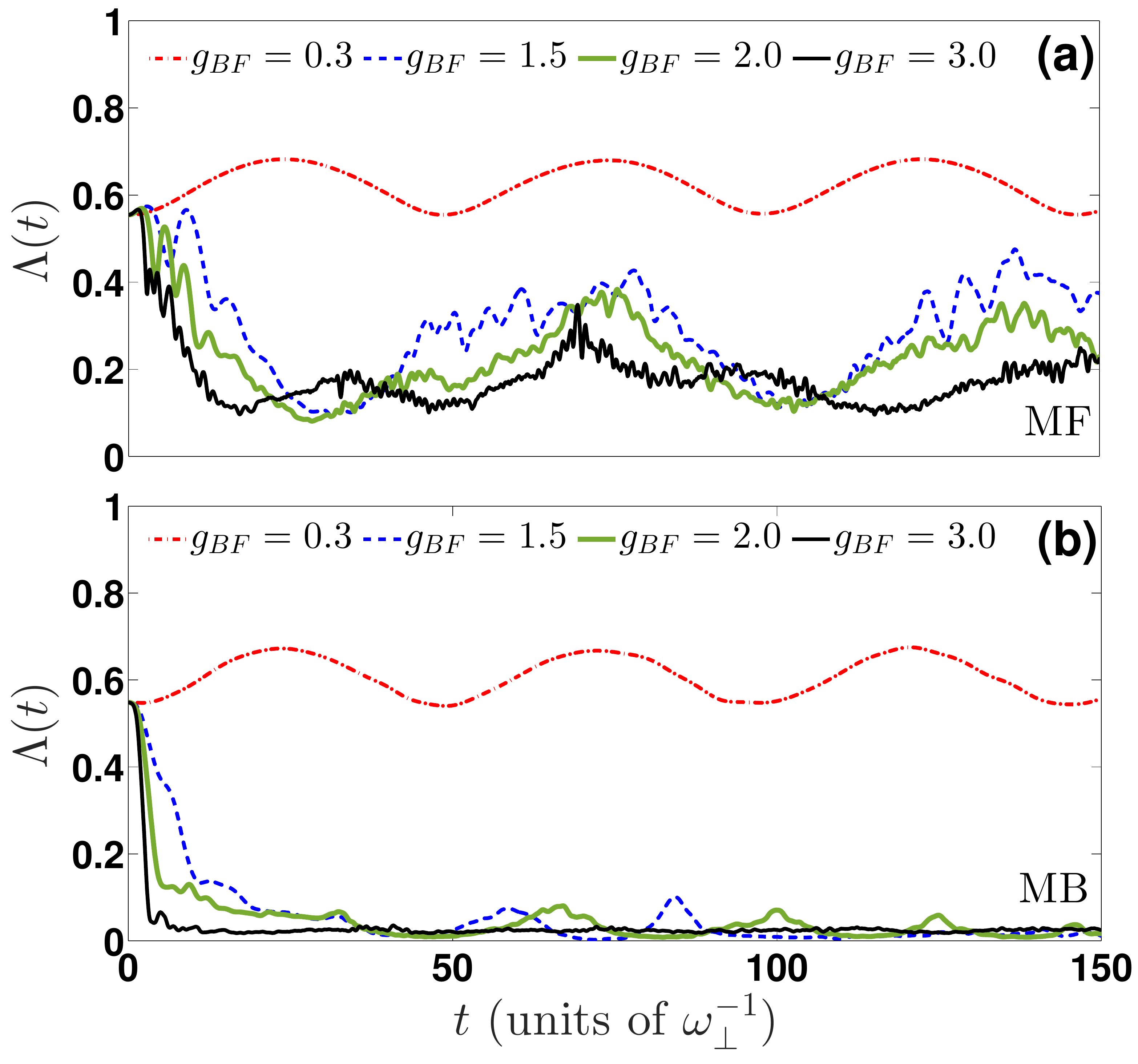}
 	\caption{Dynamics of the overlap function, $\Lambda(t)$, between the bosonic and the fermionic clouds for $g_{BB}=0.5$ and 
 	varying $g_{BF}=0.3,1.5,2,3$ (see legend) within (a) the MF and (b) the MB approach. 
 	The remaining system parameters are the same as in Fig. \ref{fig:density}.}
 	\label{fig:overlap}
\end{figure}

Quenching the interspecies interactions to very strong values, e.g. $g_{BF}=3$, we observe a significant alteration of the evolution of the BF mixture 
single-particle densities, compared to the $g_{AB}=1.5$ case, especially at the MB level [Figs. \ref{fig:density_strong} (f), (h)]. 
Evidently, within the MF approximation the formation of DB states can again be inferred. 
Notice, for instance, that the dark states are characterized by a phase jump being a multiple of $\pi$ [see the inset of Fig. \ref{fig:density_strong} (b)]. 
However, for these strong interactions the period ($T_{DB}\approx 74$) and the amplitude of the DB oscillation become larger and smaller respectively as compared 
to the $g_{BF}=1.5$ quench scenario, see Figs. \ref{fig:density_strong} (c) and (d).  
Indeed, for an increasing postquench $g_{BF}$ the degree of the dynamical phase separation between the two species is enhanced, see also the discussion in 
Sec. \ref{subsec:miscibility} and Fig. \ref{fig:overlap} (a). 
The latter dynamical process is manifested by the spontaneous generation of dark soliton structures in the bosonic gas and bright 
solitons of the fermionic component. 
Moreover, as a result of the larger degree of the dynamical phase separation for increasing $g_{BF}$ the spontaneously formed dark solitons are found to be 
deeper [e.g. see $\rho^{(1)}_B(x;t=2)$ in Figs. \ref{fig:density_strong} (a) and (b)] and thus slower, with the respective bright ones being in 
turn more spatially localized, see for instance $\rho^{(1)}_F(x;t=2)$ in Figs. \ref{fig:density_strong} (c) and (d). 
The fact that these dark structures appear to be deeper and the bright solitons more spatially localized suggests that their inverse width $D$ becomes larger. 
Indeed, measuring the inverse width $D$ of the solitons and the velocity ($\sin\phi$) of the dark component via fitting to their analytical waveforms it is found 
that for $g_{BF}=3$ only a slight increase of $D$ occurs while $\sin \phi$ decreases significantly as compared to the $g_{BF}=1.5$ case. 
Recall that \cite{Yan_DB} the velocity of a DB pair is given by $\dot{x}_0(t)=D\tan\phi$. 
However for $g_{BF}=3$ the amplitude of the dark soliton, namely $\cos \phi$, increases significantly as compared to the $g_{BF}=1.5$ 
scenario, see also Figs. \ref{fig:density_strong} (a), (b). 
As a consequence $\dot{x}_0(t)$ reduces and therefore the observed oscillations of the DB pair are of smaller amplitude. 
Also, for a DB soliton it is known \cite{Yan_DB,DB_oscil} that its oscillation period is inversely proportional to the chemical potential of the dark component 
or equivalently the Thomas-Fermi radius of the BEC background that hosts the dark states. 
Inspecting again the ground state of $\rho^{(1)}_{B}(x)$ for larger $g_{BF}$ we can deduce that the corresponding Thomas-Fermi radius becomes smaller 
for a larger $g_{BF}$ [compare Figs. \ref{fig:ground} (a)-(c)], a behavior that explains the observed increased oscillation period of the DB soliton. 
In contrast, at the MB level we observe that after the quench $\rho^{(1)}_{F}((x;t))$ splits into two branches which repel each other at the very 
early stages of the dynamics ($0<t<5$) and then for later evolution times oscillate with a very small amplitude around a mean value and tend to an almost 
steady state (see also the discussion in Secs. \ref{subsec:distance}, \ref{subsec:energy} and \ref{subsec:two_body_reduced}). 
On the other hand, the majority species forms density dips at the spatial regions where the density peaks of the impurities are located. 
Therefore, the quench generated entities resemble a two DB solitary wave state which in contrast to its MF counterpart tends to 
approach a stationary state \cite{Yan_DB,Katsimiga_DB1}. 

\subsection{Degree of miscibility}\label{subsec:miscibility}

To appreciate the degree of miscibility or immiscibility between the species of the BF mixture, 
at the one-body level we resort to the overlap integral function 
$\Lambda(t)$ \cite{mistakidis_phase_sep,Bandyopadhyay} which is defined as  
\begin{align}
\Lambda (t)=\frac{\left[\int d x \, \rho_B^{(1)}(x;t)\rho_F^{(1)}(x;t)\right]^2}{\left[\int d x\, (\rho_B^{(1)}(x;t))^2\right]\left[\int d x\, (\rho_F^{(1)}(x;t))^2\right]}.
\end{align} 
In particular, $\Lambda=0$ and $\Lambda=1$ designate zero and complete spatial overlap of the two species respectively. 
Figures \ref{fig:overlap} (a), (b) illustrate $\Lambda(t)$ for different postquench interspecies interaction strengths at the MF and the MB level respectively. 
For weak values of $g_{BF}$, such that $g_{BF}<g_{BB}$, the overlap integral oscillates in time (reflecting this way the breathing motion e.g. of the bosonic cloud) 
around the value 0.6 demonstrating a tendency towards miscibility both at the MF and the MB level. 
This behavior of $\Lambda(t)$ changes for quenches that satisfy $g_{BF}>g_{BB}$. 
Regarding the evolution within the MF approximation, independently of $g_{BF}>g_{BB}$, $\Lambda(t)$ shows a decreasing behavior at the very early stages of the dynamics and subsequently performs 
oscillations [resembling this way the overall motion of the DB state, see Figs. \ref{fig:density_strong} (a), (c)] around a mean value which is smaller for a larger 
postquench $g_{BF}$, e.g. it is 0.22 for $g_{BF}=1.5$ and 0.18 for $g_{BF}=3$.  
Accordingly, also the amplitude of the oscillations of $\Lambda(t)$ becomes smaller for increasing $g_{BF}$ indicating an overall tendency for a larger degree of phase separation 
for stronger postquench interactions. 
It is also worth noticing at this point that $\Lambda(t)\neq0$ even for $g_{BF}=3$, thus evincing that complete phase separation can not be 
achieved at the MF level. 
However, at the MB level $\Lambda(t)$ shows a completely different behavior and its magnitude is always smaller than its MF counterpart, in particular compare 
Figs. \ref{fig:overlap} (a) and (b) for a fixed $g_{BF}>g_{BB}$.  
Indeed, entering the strong interspecies interaction regime, $g_{BF}>1.5$, we observe that $\Lambda(t)$ exhibits a fast decrease and then for later evolution times 
approaches zero thus testifying a complete phase separation between the two species.

\subsection{Relative distance between the impurities}\label{subsec:distance}

\begin{figure}[ht]
 	\includegraphics[width=0.48\textwidth]{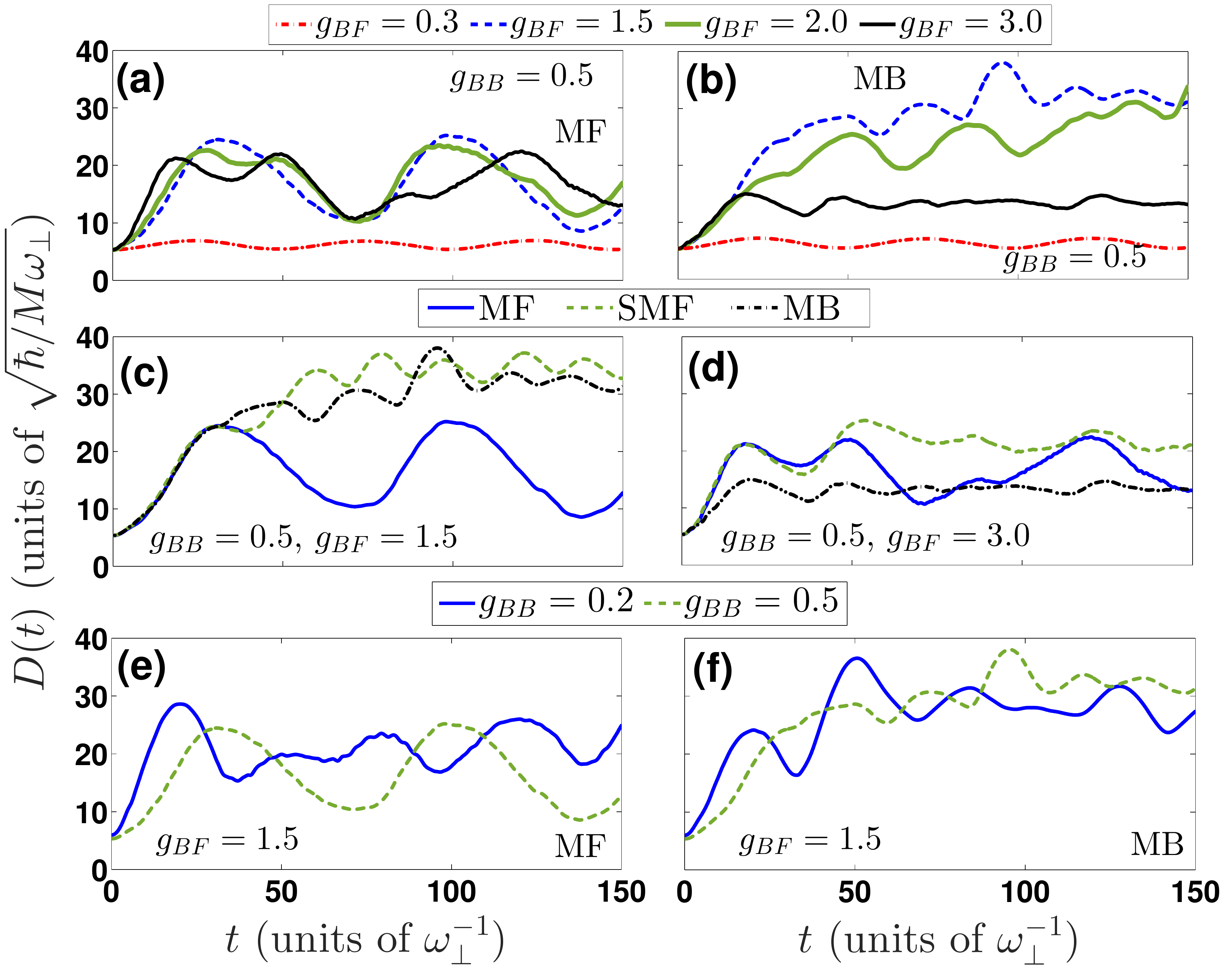}
 	\caption{Time-evolution of the relative distance, $D(t)$, between the two fermionic impurities following an interspecies interaction quench of the 
 	BF mixture. 
 	(a), (b) shows $D(t)$ for fixed $g_{BB}=0.5$ and varying $g_{BF}= 0.3,1.5, 2, 3$ calculated within (a) the MF and (b) the MB approach. 
 	(c), (d) illustrates $D(t)$ for $g_{BB} = 0.5$ and for (c) $g_{BF}=1.5$ and (d) $g_{BF}=3.0$ within different levels of approximation (see legend).
 	(e), (f) presents $D(t)$ for $g_{BF} = 1.5$ and varying $g_{BB} = 0.2, 0.5$ (see legend) at (e) the MF and (f) the MB level. 
 	Note that the Thomas-Fermi radius of the bosonic gas corresponds to $R_{TF}=14.4$ at $g_{BB}=0.2$ and $R_{TF}=19.6$ at $g_{BB}=0.5$. 
 	The other system parameters are the same as in Fig. \ref{fig:density}.}
 	\label{fig:distance} 
 \end{figure}

To estimate the nature of the interactions induced by the presence of the bosons between the two non-interacting fermionic impurities we determine their relative distance, $D(t)$, 
during the dynamics [see also Eq. (\ref{eq:distance})]. 
Recall, that this quantity can be probed experimentally via in-situ spin-resolved single-shot measurements performed on the state of the impurities \cite{Jochim2}. 
The time-evolution of $D(t)$ is shown in Figs. \ref{fig:distance} (a), (b) for different $g_{BF}$ within the MF and the MB approach respectively. 
For very weak interactions, e.g. $g_{BF}=0.3$, and in both approaches $D(t)$ takes small values since the impurities are very close in this case and undergoes small amplitude 
oscillations in time reflecting this way the breathing motion of the impurities. 
Focusing on the MF approximation we observe that for every $g_{BF}\gg g_{BB}$ the relative distance exhibits an increasing tendency at the initial stages of the dynamics while 
for later times oscillates with a decreasing amplitude for larger $g_{BF}$. 
The initially increasing tendency of $D(t)$ essentially demonstrates the repulsive tendency of the impurity density branches already evident in $\rho^{(1)}_{F}((x;t))$ 
[e.g. see Fig. \ref{fig:density_strong} (c)], which is a consequence of the effective potential created by $\rho^{(1)}_B$. 
The subsequent fluctuating behavior of $D(t)$ reflects the oscillatory motion of the previously discussed bright states. 
However at the MB level the behavior of $D(t)$ changes drastically. 
In particular for a quench to $g_{BF}\gg g_{BB}$, $D(t)$ initially increases reflecting the presence of repulsive induced interactions captured by $\rho^{(1)}_{F}((x;t))$ 
[see also Fig. \ref{fig:density_strong} (g)] while small amplitude oscillations occur during evolution, with an overall increasing tendency. 
For very strong postquench interaction strengths, e.g. for $g_{BF}=3$, the amplitude of these oscillations diminishes and $D(t)$ acquires an almost constant 
value [see also Fig. \ref{fig:density_strong} (h)] suggesting that the impurities tend to approach a stationary state.   

\begin{figure}[ht]
 	\includegraphics[width=0.46\textwidth]{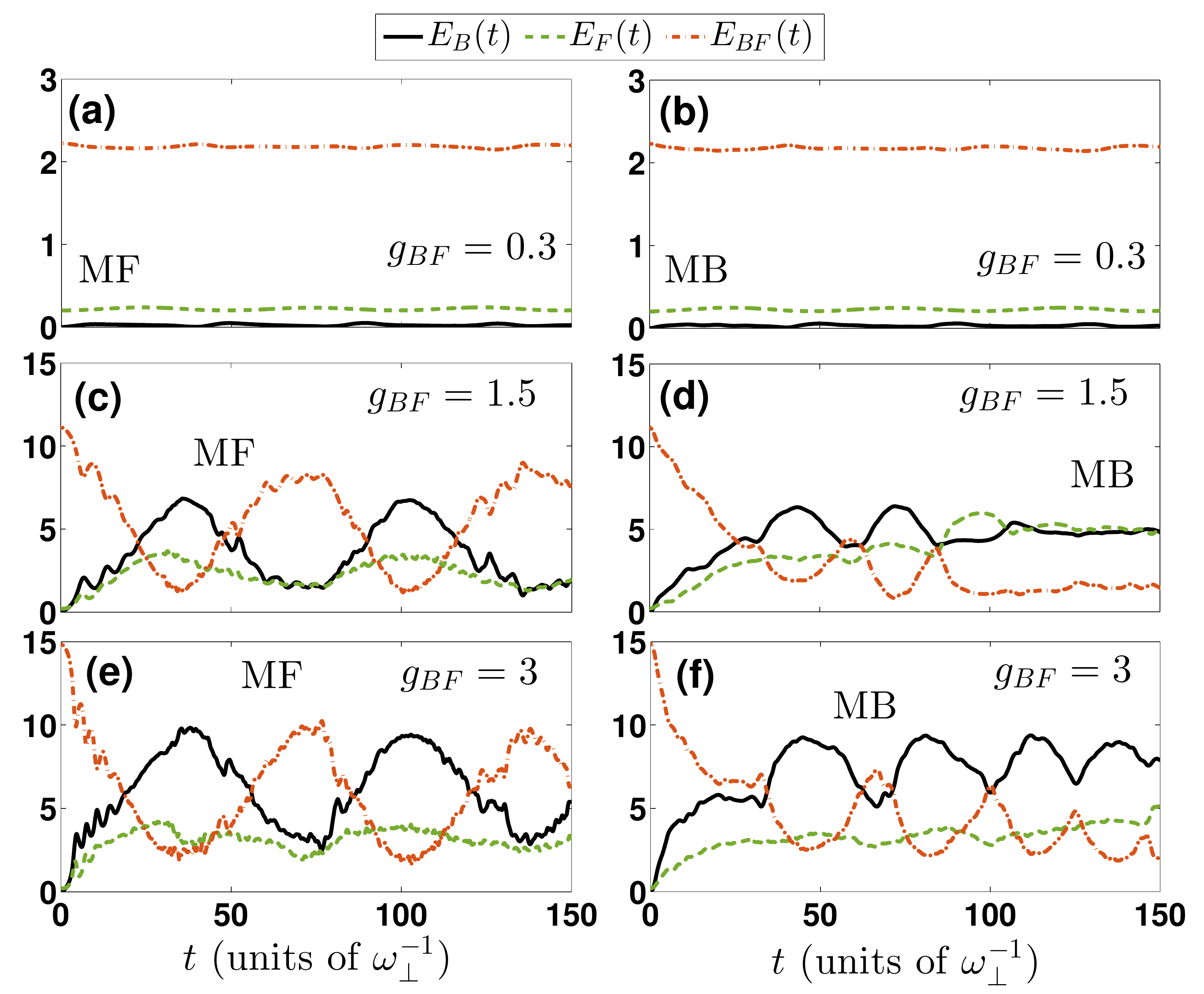}
 	\caption{Expectation value of the energy of the bosons $E_{B}(t)$, the fermions $E_F(t)$ and their mutual interaction $E_{BF}(t)$ (see legend) 
 	following an interspecies interaction quench from $g_{BF}=0.1$ to (a), (b) $g_{BF}=0.3$, (c), (d) $g_{BF}=1.5$ and (e), (f) $g_{BF}=3$.  
 	The calculations are performed within (a), (c), (e) the MF approximation and (b), (d), (f) the MB approach. 
 	The other system parameters are the same as in Fig. \ref{fig:density}.}
 	\label{fig:energy} 
 \end{figure}

To expose the significance of the inclusion of different levels of correlations in the dynamical behavior of the relative distance we next measure $D(t)$ 
between the fermionic impurities by systematically taking into account different orders of correlations. 
In particular, we determine $D(t)$ within the MF approximation where all correlations are neglected [see Eq. (\ref{Eq:MF})], the SMF approximation where 
only intraspecies correlations are included [see Eq. (\ref{eq:SMF})] and in the MB approach where all correlations are incorporated [see Eq. (\ref{Eq:WF})]. 
Figures \ref{fig:distance} (c), (d) present $D(t)$ within the above-mentioned approaches for $g_{BF}=1.5$ and $g_{BF}=3$ respectively. 
As it can be seen, for $g_{BF}=1.5$ all approaches capture the initial ($0<t<35$) increase of $D(t)$ but for later times significant deviations occur. 
Indeed, at the MF level $D(t)$ undergoes large amplitude oscillations in time, while at the SMF and MB level $D(t)$ shows an overall increasing tendency performing 
also small amplitude oscillations. 
Importantly here the SMF approximately captures the overall behavior of $D(t)$ but overestimates its values. 
Turning to stronger interactions, $g_{BF}=3$ we observe that both the MF and the SMF fail to capture the MB dynamics of $D(t)$ [Fig. \ref{fig:distance} (d)]. 
Concluding we can clearly infer that both intra- and interspecies correlations are important for the adequate description of $D(t)$. 
The latter in turn disctates a much stronger repulsion between the impurities for intermediate $g_{BF}$ values when compared to a MF description of the 
out-of-equilibrium dynamics. 

The behavior of $D(t)$ for different intraspecies interaction strengths, $g_{BB}$, of the bosonic medium but for fixed $g_{BF}=1.5$ is illustrated in 
Figs. \ref{fig:distance} (e), (f) within the MF and the MB approach respectively. 
In both cases and for all $g_{BB}$ values that we have addressed it is found that at initial times $D(t)$ increases and consecutively oscillates 
as time evolves. 
However, this initial increase occurs faster for weaker intraspecies interactions. 
In particular, $D(t)$ reaches a local maximum with $D(t\approx20)\approx29$ and $D(t\approx20)\approx23$ within the MF and the MB approach respectively 
for $g_{BB}=0.2$. 
In contrast to the above and e.g. for $g_{BB}=0.5$, the first maximum appears for comparatively larger times. 
Here, $D(t\approx31)\approx25$ and $D(t\approx50)\approx29$ for the MF and the MB case respectively. 
It is worth mentioning that a larger $g_{BB}$ implies also a larger Thomas-Fermi radius, and hence size, of the BEC. 
For instance, the Thomas-Fermi radius at $g_{BB}=0.2$ is $R_{TF}\approx14$ while at $g_{BB}=0.5$ becomes $R_{TF}\approx19$. 
Therefore, the observed differences in the initial increase and subsequent oscillations of $D(t)$ for varying $g_{BB}$ can be attributed to the fact 
that a larger $g_{BB}$, while fixing the particle number $N_B$, gives rise to a more dilute repulsive environment. 
The latter in turn exerts a weaker repulsive effective force on the impurities compared to the case of a smaller $g_{BB}$. 
Thus, a stronger $g_{BB}$ slows down the initial expansion of the impurities and vice versa. 
The latter effect takes place at both the MF and the MB level. 
Moreover, for evolution times $t>80$ almost damped oscillations of $D(t)$ for all $g_{BB}$ values occur when correlations are present. 
A behavior that is absent within the MF treatment (see also the discussion below).

\subsection{Interspecies energy transfer}\label{subsec:energy} 

To further analyze and understand the nonequilibrium dynamics of the BF mixture occurring for different postquench interspecies interaction strengths  
below we focus on the study of the distinct energy contributions \cite{Mistakidis_orthog_catastr,Mistakidis_coh_state}. 
More specifically, the normalized energy of the bosonic species is $E_B(t)=\braket{\Psi(t)|\hat{T}_B+\hat{V}(x)+\hat{H}_{BB}|\Psi(t)}-\braket{\Psi(0)|\hat{T}_B+\hat{V}(x)+\hat{H}_{BB}|\Psi(0)}$, 
and for the fermionic impurities corresponds to $E_F(t)=\braket{\Psi(t)|\hat{T}_F+\hat{V}(x)|\Psi(t)}$. 
Additionally, the interspecies interaction 
energy is $E_{BF}(t)=\braket{\Psi(t)|\hat{H}_{BF}|\Psi(t)}$. 
Note that the kinetic and the potential energy operators of the $\sigma=B,F$ species are 
$\hat{T}_{\sigma}=-\int dx \hat{\Psi}^{\sigma \dagger}(x)\frac{\hbar^2}{2 M} (\frac{d}{dx^{\sigma}})^2 \hat{\Psi}^{\sigma}(x)$ and 
$\hat{V}_{\sigma}=\int dx \hat{\Psi}^{\sigma \dagger}(x)\frac{1}{2} M \omega^2 x^2 \hat{\Psi}^{\sigma}(x)$ respectively. 
The operators of the intra- and interspecies interactions read $\hat{H}_{BB}=g_{BB} \int dx~\hat{\Psi}^{B \dagger}(x) \hat{\Psi}^{B \dagger}(x) \hat{\Psi}^{B} (x)\hat{\Psi}^{B}
(x)$ and $\hat{H}_{BF}=g_{BF}\int dx~\hat{\Psi}^{B \dagger}(x) \hat{\Psi}^{F \dagger}(x) \hat{\Psi}^{F}(x)\hat{\Psi}^{B}(x)$. 
Also, $\hat{\Psi}^{\sigma} (x)$ [$\hat{\Psi}^{\sigma \dagger} (x)$] denotes the $\sigma$ species field operator that annihilates [creates] a $\sigma$ species particle at position $x$.  

The time-evolution of the above-mentioned energy contributions is shown in Fig. \ref{fig:energy} for a varying $g_{BF}$ both within the 
MF approximation [Figs. \ref{fig:energy} (a), (c), (e)] and in the MB approach [Figs. \ref{fig:energy} (b), (d), (f)]. 
Referring to weak postquench interactions, e.g. $g_{BF}=0.3$, all energy parts are mainly constant throughout the dynamics and $E_{B}(t)<E_{F}(t)<E_{BF}(t)$ holds. 
This result being almost identical in both the MF and the MB scenario [Figs. \ref{fig:energy} (a), (b)] suggests that in the weak interaction regime the role of correlations 
is negligible. 
Turning to stronger interactions, e.g. $g_{BF}=1.5$, we observe that the different energy contributions undergo a much more involved dynamics 
in both approaches [Figs. \ref{fig:energy} (c), (d)]. 
Indeed, when all particle correlations are neglected $E_{BF}(t)$ decreases while $E_{B}(t)$ and $E_{F}(t)$ increase at the initial stages of the dynamics ($0<t<35$). 
For later times all the different energy contributions exhibit large amplitude oscillations in time being more pronounced for $E_{B}(t)$. 
In particular, $E_{B}(t)$ and $E_{F}(t)$ oscillate in-phase with respect to one another and both are out-of-phase with $E_{BF}(t)$. 
This latter behavior suggests that a periodic energy transfer process from the fermionic impurities (bright solitons) to the bosonic environment occurs as a consequence of 
the DB oscillatory motion. 
More specifically, when the bright solitons travel towards the edges of the BEC medium [see also Fig. \ref{fig:density_strong} (c)] they acquire more kinetic energy, 
and thus $E_{F}(t)$ increases, and convey energy to the bosonic bath resulting to an increase of the energy, $E_{B}(t)$, of the latter. 
This energy stems from the large $E_{BF}$ at $t=0$ which subsequently decreases. 
On the other hand, when the bright solitons are reflected back to the trap center [Fig. \ref{fig:density_strong} (c)] they become slower \cite{Katsimiga_DB1,Yan_DB} 
and therefore $E_{F}(t)$ and consequently $E_{B}(t)$ become smaller while $E_{BF}(t)$ increases since the interaction between the two species is larger.    
However, the dynamical behavior of the corresponding energy contributions within the correlated treatment is drastically different when compared to its 
MF counterpart [Figs. \ref{fig:energy} (c) and (d)]. 
Recall that such a deviation is already evident from the corresponding single-particle density evolution, see e.g. Figs. \ref{fig:density_strong} (c) and (g). 
Initially, $0<t<35$, $E_{BF}(t)$ reduces while $E_{F}(t)$ and $E_{B}$ increase. 
Indeed, within this time interval the two fermionic density branches move to the edges of the bosonic bath [Fig. \ref{fig:density_strong} (g)] with a large kinetic 
energy and as a result devolve energy to the latter. 
For $35<t<95$, $E_{B}(t)$ and $E_{BF}(t)$ oscillate out-of-phase whilst $E_{F}(t)$ increases while performing small amplitude oscillations.  
Therefore the fermionic impurities transfer a part of their energy to the bosonic gas. 
It is in this time interval that $\rho^{(1)}_{F}(x;t)$ resides close to the boundaries of $\rho^{(1)}_{B}(x;t)$ and still weakly interacts with the BEC. 
Deeper in the evolution, $t>95$, all energy components acquire an almost constant value with $E_{B}(t)\equiv E_{F}(t)<E_{BF}(t)$. 
Recall that for $t>95$ $\rho^{(1)}_{F}(x;t)$ resides at the edges of $\rho^{(1)}_{B}(x;t)$ and therefore the two species barely interact. 

Inspecting the energies for even stronger interactions, $g_{BF}=3$, [Figs. \ref{fig:energy} (e), (f)] we can deduce that an overall similar to the 
above-described dynamics takes place. 
Focusing on the MF approximation we observe that at the initial time period ($0<t<30$) the decrease of $E_{BF}(t)$ is accompanied by a simultaneous increase 
of both $E_{F}(t)$ and $E_{B}(t)$ with the magnitude of the latter found to be larger. 
We remark that in this time interval the bright solitons move to the edges of the BEC cloud thus becoming faster [Fig. \ref{fig:density_strong} (d)] and transfering energy to the bosonic bath. 
For $t>30$, $E_{F}(t)$ and $E_{B}(t)$ oscillate in-phase with each other but out-of-phase with $E_{BF}(t)$. 
Note also the much larger oscillation amplitude of $E_{B}(t)$ as compared to $E_{F}(t)$. 
Overall when the bright solitons travel to the edges (core) of the BEC medium they acquire more (less) kinetic energy. 
The corresponding $E_{BF}(t)$ becomes smaller (larger), and the bosonic bath gains (looses) energy [see also Figs. \ref{fig:density_strong} (b), (d)].  
On the contrary, turning to the MB description the energies show again a quite different behavior to the one observed in the MF approximation. 
For times up to $t=10$, $E_{BF}(t)$ decreases favoring an increase of $E_{F}(t)$ and $E_{B}(t)$ such that $E_{F}(t)\ll E_{B}(t)$. 
Recall that in this time interval $\rho^{(1)}_{F}(x;t)$ breaks into two repelling density branches [Fig. \ref{fig:density_strong} (h)]. 
As a consequence the increase of $E_{F}(t)$ can be attributed to the increasing kinetic energy of the fermions, in this time interval, that leads to a transfer of energy 
from the fermions to the bosons. 
For $t>10$, $E_{B}(t)$ exhibits an overall increasing tendency oscillating out-of-phase with $E_{BF}(t)$ which in turn decreases. 
Simultaneously, $E_{F}(t)$ slightly increases in time. 
This approximately constant behavior of $E_{F}(t)$ essentially reflects the almost steady behavior of $\rho^{(1)}_{F}(x;t)$ during evolution [see also Fig. \ref{fig:density_strong} (h)]. 
Another important observation here is that $E_{B}(t)$ is found to be larger for $g_{BF}=3$ when compared to the relevant energy contribution for $g_{BF}=1.5$, 
indicating that for these stronger interspecies interactions the energy gain of the bosonic bath is even larger.

\subsection{Degree of correlations}\label{subsec:degree_cor}
 
To quantify the correlated nature of the BF mixture dynamics we next estimate the degree of entanglement (interspecies correlations) by employing $S_{VN}(t)$ [Eq. (\ref{eq:entropy})]. 
Additionally, we invoke $F_{\sigma}(t)$ [Eq. (\ref{fragmentation})] to infer about the fragmentation (intraspecies correlations) of the system's MB state [Eqs. (\ref{Eq:WF}) and (\ref{Eq:SPF})]. 
Recall that $S_{VN}(t)\neq0$ designates the presence of entanglement otherwise the system is termed non-entangled \cite{mistakidis_phase_sep,darkbright_beyond}. 
Also, when $F_{\sigma}(t)>0$ the $\sigma$-species is said to be intraspecies correlated (see also Sec. \ref{observables}). 
Most importantly, since the fermionic species consist of spin-polarized, namely non-interacting, fermions the existence of their intraspecies correlations 
during evolution is caused by the presence of the interspecies correlations.
\begin{figure}[ht]
  	\includegraphics[width=0.45\textwidth]{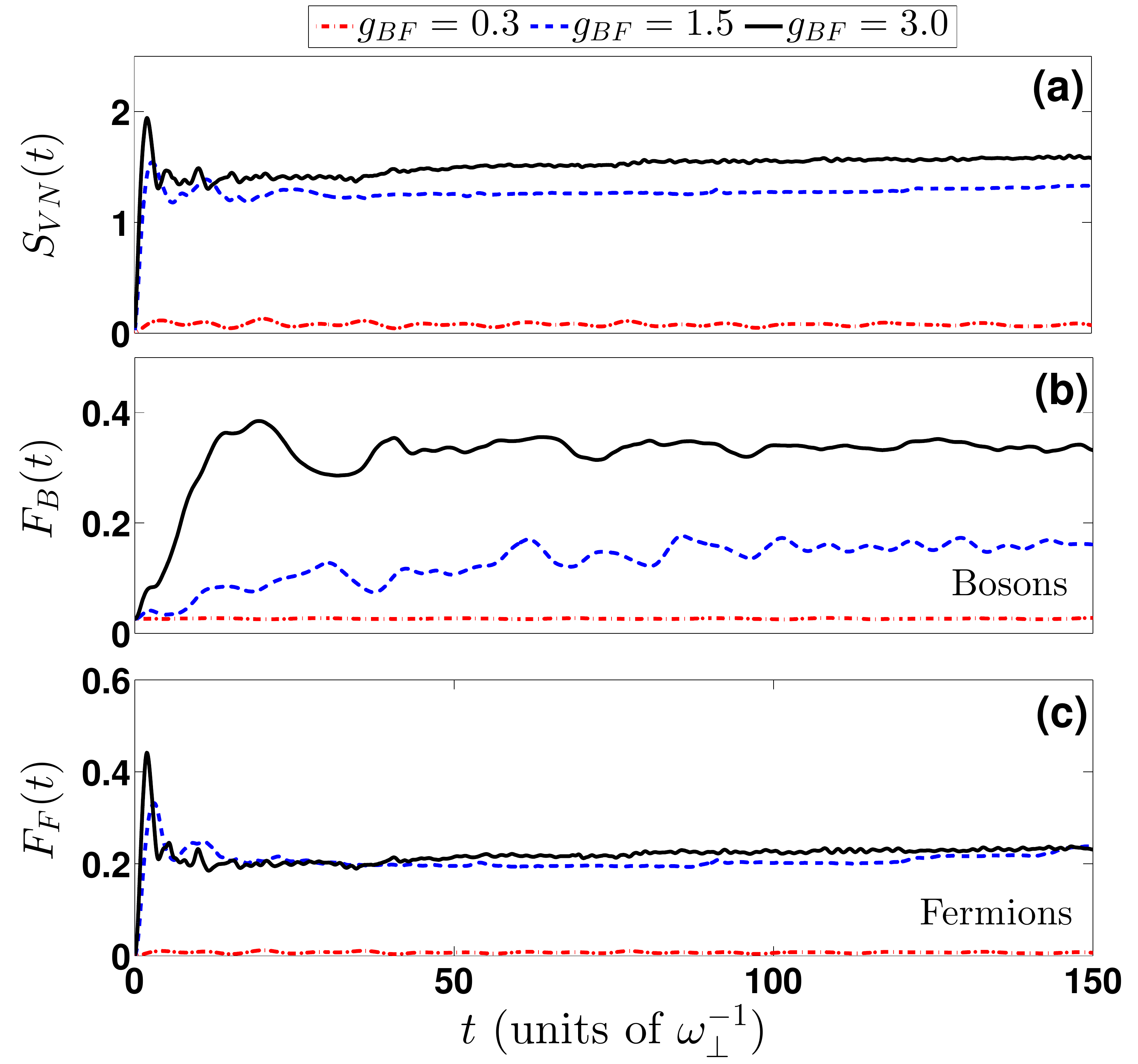}
  	\caption{(a) Von-Neumann entropy for fixed $g_{BB}=0.5$ and different postquench $g_{BF}$ (see legend). 
  	(b) Deviation from unity of the first natural population of the bosonic cloud for $g_{BB}=0.5$ and varying 
  	$g_{BF}$. 
  	(c) Deviation from unity of the first two natural populations of the fermions for $g_{BB}=0.5$ and 
  	distinct $g_{BF}$ (see legend). 
  	Other system parameters are the same as in Fig. \ref{fig:density}.}
  	\label{fig:vonNeumann}
\end{figure}

Figure \ref{fig:vonNeumann} presents $S_{VN}(t)$ and $F_{\sigma}(t)$ during the interspecies interaction quench dynamics of the BF mixture for a varying $g_{BF}$. 
Recall that the initial (ground state) single-particle densities of the $\sigma$ species [see the discussion in Sec.  \ref{sec:ground_state}] are almost identical 
within the MF and the MB description [Figs. \ref{fig:ground} (a), (d)]. 
As a result at $t=0$ the BF mixture is mainly uncorrelated and this is also confirmed by the fact that $S_{VN}(0)\approx0$, $F_{F}(0)\approx 0$ 
and $F_{B}(0)\approx 0.02$ [see Fig. \ref{fig:vonNeumann}]. 
On the other hand, inspecting the time-evolution of the above-mentioned quantities we can infer about the presence of both inter- and 
interspecies correlations. 
For weak postquench interactions, $g_{BF}=0.3$, there is only a small amount of inter- and intraspecies correlations since both 
$S_{VN}(t)$ and $F_{\sigma}(t)$ are suppressed taking very small values and being almost constant in time. 
However, for stronger interactions such as $g_{BF}=1.5,3$ the entropy $S_{VN}(t)$ as well as $F_{\sigma}(t)$ increase during 
the evolution and tend to saturate to a certain finite value. 
This indicates that the underlying MB state is strongly both entangled and fragmented. 
Also stronger postquench effective interspecies interactions, $g_{BF}$, result to larger values of $S_{VN}(t)$ and $F_{\sigma}(t)$. 
Another interesting observation is that $S_{VN}(t)$ and $F_{F}(t)$ increase more rapidly at the initial stages of the dynamics 
than $F_{B}(t)$ which grows in a slower manner. 
This indicates that the interspecies correlations and fermionic intraspecies correlations are stronger than the bosonic intraspecies 
ones (see also below). 

Moreover, we can infer that for strong postquench interactions, e.g. $g_{BF}=1.5$, all the above-described correlation measures show a tendency 
to saturation for times $t>80$. 
Indeed by inspecting the dynamics of the corresponding natural species populations $\lambda_k(t)$ with $k=1,2,\dots,10$ as well as the natural populations of the 
orbitals of each species, namely $n_i^B(t)$ ($i=1,2,3$) and $n_i^F(t)$ ($i=1,2,\dots,8$), we observe that they rapidly oscillate for $t<80$ while at later times a slower small amplitude 
redistribution takes place (results not shown here for brevity). 
This slow redistribution of $\lambda_k(t)$, $n_i^B(t)$, and $n_i^F(t)$ is essentially imprinted in $S_{VN}(t)$, $F_{B}(t)$ and $F_{F}(t)$ respectively and it is reminiscent 
of a prethermalization phenomenon \cite{thermalization} occuring at the individual correlation levels \cite{thermal_axel,thermal_bera}. 
Recall that such a saturation behavior occurs also for other observables, e.g. the relative distance between the two fermions for $g_{BF}=3$ [Fig. \ref{fig:distance}], further 
supporting the tendency of the system to prethermalize. 
However in order to strictly infer about the occurence of prethermalization dynamics requires a further investigation and an in-depth analysis. 
More specifically, one needs to compare e.g. with the predictions of the corresponding Gaussian orthogonal ensemble 
of random matrices, and also monitor the time-evolution of the system for longer evolution times, an investigation which certainly lies 
beyond the scope of the present effort. 

\begin{figure}[ht]
  	\includegraphics[width=0.5\textwidth]{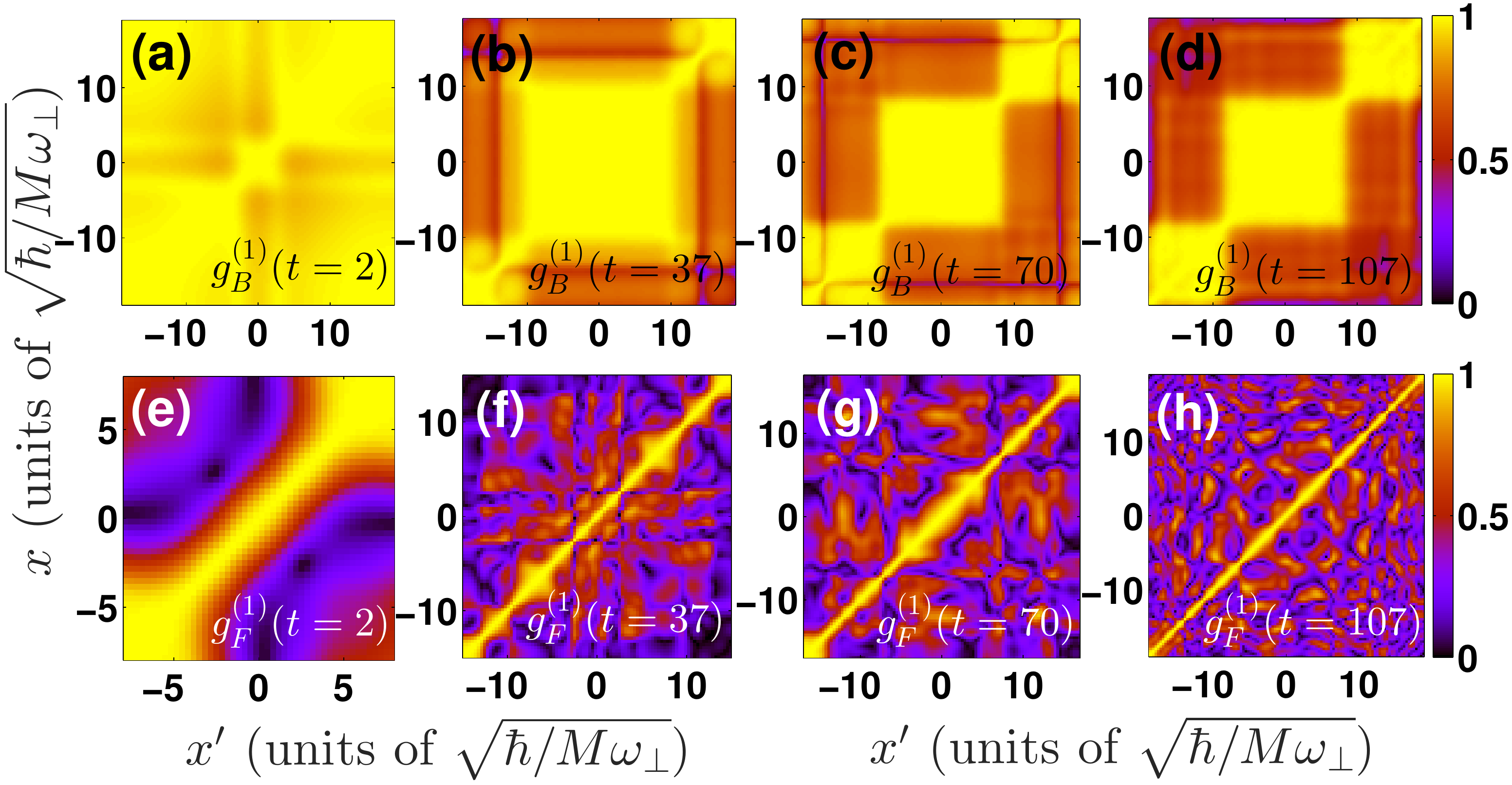}
  	\caption{(a)-(d) Snapshots of the one-body coherence function at different time instants (see legends) of the evolution following an 
  	interspecies interaction quench from $g_{BF}=0.1$ to $g_{BF}=1.5$. 
  	(e)-(h) The same as before but for the fermions.  
  	The remaining system parameters are the same as in Fig. \ref{fig:density_strong}.}
  	\label{fig:coherence}
\end{figure}

\begin{figure}[ht]
  	\includegraphics[width=0.48\textwidth]{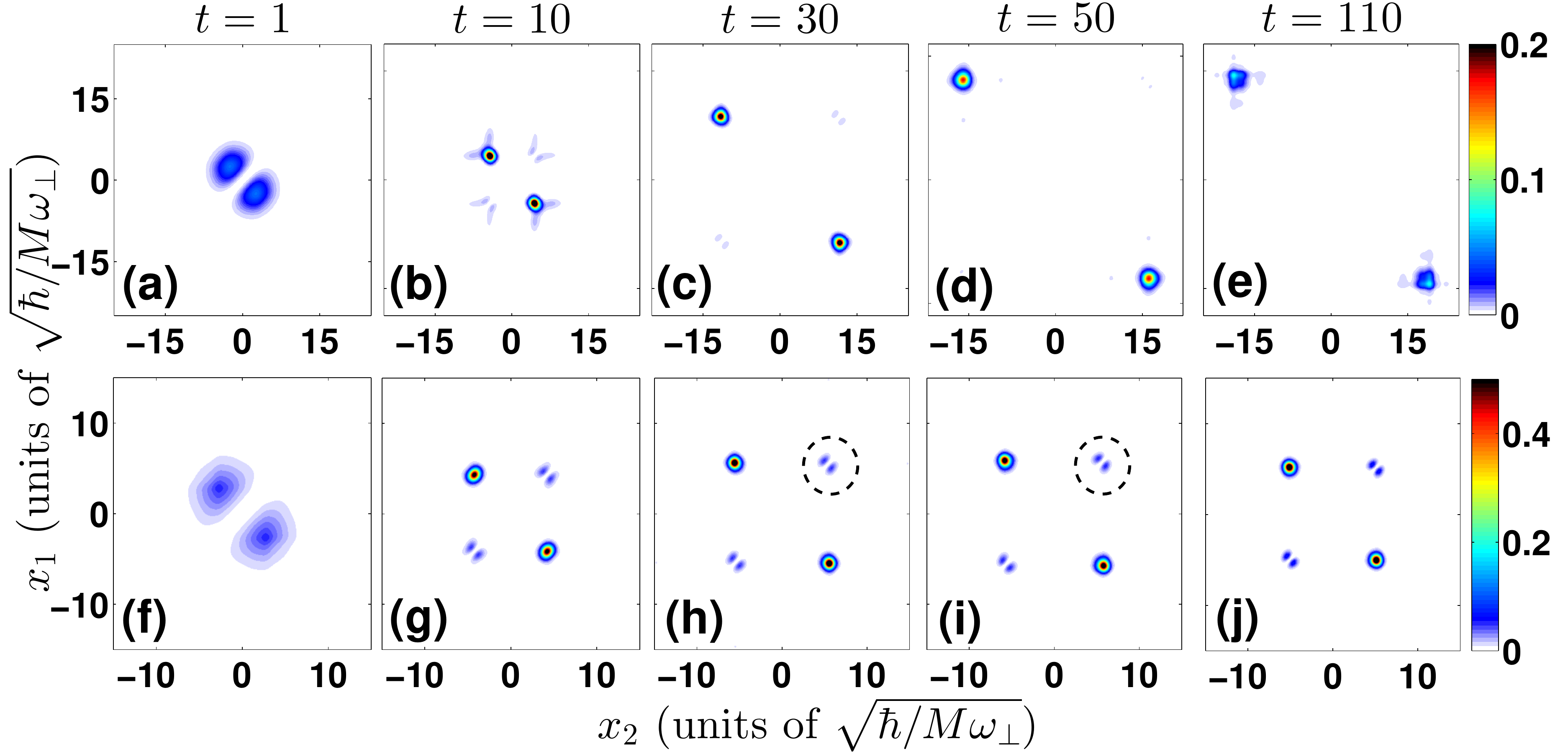}
  	\caption{(a)-(d) The fermionic two-body reduced density matrix at different time instants (see legends) of the evolution 
  	upon considering an interspecies interaction quench from $g_{BF}=0.1$ to (a)-(e) $g_{BF}=1.5$ and (f)-(j) $g_{BF}=3$. 
  	The dashed circles in (h), (i) indicate the population close to the diagonal elements for $x_1, x_2>0$. 
  	The remaining system parameters are the same as in Fig. \ref{fig:density_strong}.}
  	\label{fig:two_body}
\end{figure}

\subsection{Evolution of the one-body coherence}\label{subsec:coherence}

In order to further expose the role of correlations in the above-discussed interaction quench dynamics of the BF mixture we finally resort to 
the corresponding $\sigma$ species one-body coherence function $g_{\sigma}^{(1)}(x,x';t)$ of Eq. (\ref{one_body_cor}) [see Fig. \ref{fig:coherence}]. 
Closely inspecting $g_B^{(1)}(x,x^\prime;t)$ we can deduce that the bosonic species is fully coherent within the time interval ($0<t<50$) 
that the two fermionic density branches travel towards its edges [see also Figs. \ref{fig:density_strong} (e), (g)]. 
Indeed, for this time interval $g_B^{(1)}(x,x';t)\approx1$ for every $x$, $x'$ lying within the spatial extent of the bosonic cloud [Figs. \ref{fig:coherence} (a), (b)]. 
For later evolution times where the aforementioned fermionic density branches probe the edges of the bosonic species we observe that losses of coherence, 
namely $g_B^{(1)}(x,x';t)<1$, occur either between the two edges, e.g. $10<x'<15$ and $-15<x<-10$, or between one edge and the core of the bosonic cloud, e.g. $-8<x'<8$ and $10<x<15$. 
Regarding the coherence of the impurity species we can infer that each fermionic density branch is perfectly coherent with itself having $g_A^{(1)}(x,x'=x;t)\approx1$ 
[Figs. \ref{fig:coherence} (e)-(h)] but it is almost fully incoherent with the other branch, since e.g. $g_A^{(1)}(x=8,x'=-8;t)\approx0.3$ throughout the 
dynamics [Figs. \ref{fig:coherence} (g)-(h)]. 
Therefore we can conclude that during the evolution the BF mixture develops Mott-like one-body correlations \cite{Sherson,darkbright_beyond,mistakidis_phase_sep} that emerge 
between the two fermionic density branches indicating their spatial localization.   

\subsection{Dynamics of the two-body reduced density matrix}\label{subsec:two_body_reduced}

Another interesting prospect is to unravel the spatially resolved dynamics of the two fermionic impurities with respect to one another. 
Note that the impurities are dressed by the excitations of the bosonic gas forming quasiparticles. 
In turn, these quasiparticles can move independently or interact thereby possibly forming a bound state in the latter case. 
To track the underlying two particle dynamics we invoke 
the time-evolution of the corresponding diagonal of the two-body fermionic reduced density matrix 
\begin{equation}
\begin{split}
\rho^{(2)}_{FF}(x_1,x_2;t)=\bra{\Psi_{MB}(t)}\hat{\Psi}^{F \dagger}(x_1) \hat{\Psi}^{F \dagger}(x_2)\\ \times \hat{\Psi}^{F}(x_1) \hat{\Psi}^{F}(x_2)\ket{\Psi_{MB}(t)}. 
\end{split}
\end{equation}
Here, $\hat{\Psi}^{F}(x_1)$ denotes the fermionic field operator that annihilates a fermion at position $x_1$. 
Recall that the diagonal of the two-body reduced density matrix provides the probability of measuring simultaneously one fermion at location $x_1$ and the 
other at $x_2$ \cite{mistakidis_phase_sep,Erdmann_phase_sep,Erdmann_tunel}. 
The dynamics of $\rho^{(2)}_{FF}(x_1,x_2;t)$ at specific time-instants is presented in Fig. \ref{fig:two_body} when following an interspecies 
interaction quench from $g_{BF}=0.1$ to $g_{BF}=1.5$ [Figs. \ref{fig:two_body} (a)-(e)] and $g_{BF}=3$ [Figs. \ref{fig:two_body} (f)-(j)]. 
Referring to $g_{BF}=1.5$, we observe that initially [Fig. \ref{fig:two_body} (a)] the two fermions reside close to the trap center with one   
of them located in the vicinity of $0<x_1<5$ and the other at $-5<x_2<0$ as shown by the populated anti-diagonal elements of $\rho^{(2)}_{FF}(x_1,x_2;t)$. 
The existence of a correlation hole ocurring in the diagonal of $\rho^{(2)}_{FF}(x_1,x_2=x_1;t)$ throughout the evolution is, of course, a consequence 
of Pauli's exclusion principle. 
At the early stages of the dynamics [Fig. \ref{fig:two_body} (b)] the two fermions move in the opposite direction with respect to one another, while as time 
evolves their distance becomes larger, see Figs. \ref{fig:two_body} (c)-(e). 
This increase of their relative distance has already been discussed in Section  \ref{subsec:distance} and visualized in Fig. \ref{fig:distance} (b).  
Therefore each fermion resides in one of the density branches emerging in $\rho^{(1)}_{F}(x;t)$ [Fig. \ref{fig:density_strong} (g)]. 
As a consequence we can infer the formation of two largely independent quasiparticles. 
Note here that these quasiparticles are located well inside the bosonic gas until $t\approx60$, while for later times reside at the edges of the Thomas-Fermi radius 
of the BEC [see also Fig. \ref{fig:density_strong} (g)] and are held at this location by the external harmonic trap. 
Such an independent fermionic quasiparticle formation has already been reported in \cite{Dehkharghani_PRL}. 

Turning to very strong postquench interactions, i.e. $g_{BF}=3$, we observe an overall same phenomenology as before but most importantly in this case 
there is also a non-negligible population of the $\rho^{(2)}_{FF}(x_1,x_2;t)$ elements close to its diagonal, e.g. see the dashed circles in Figs. \ref{fig:two_body} (h), (i). 
The latter contribution indicates an additional emergent attractive tendency between the two fermions and it is suggestive of the formation of a bound state between them. 
In particular, at short evolution times [Fig. \ref{fig:two_body} (f)] the two fermions are well separated residing one at $0<x_1<5$ and the other at $-5<x_2<0$, 
see the anti-diagonal elements of $\rho^{(2)}_{FF}(x_1,x_2;t)$. 
As time evolves we observe that the two fermions can be essentially in two different two-body configurations (see below). 
Namely, either the two fermions remain spatially separated throughout the dynamics [see the anti-diagonal elements of $\rho^{(2)}_{FF}(x_1,x_2;t)$] or they are very close to each other 
[see the diagonal elements of $\rho^{(2)}_{FF}(x_1,x_2;t)$] which is reminiscent of the formation of a bound state between them. 
Notice also that the occurrence of this latter two-body configuration is also responsible for the emergent attraction between the two density branches 
appearing in the corresponding fermionic single-particle density evolution [Fig. \ref{fig:density_strong} (h)]. 
Similar bound state formation has been discussed in the context of bosonic impurities \cite{Dehkharghani_PRL,Klein,Camacho_Guardian}. 
The appearance of these two different two-body configurations can be understood by resorting to the spectral decomposition of the two-body density matrix. 
The latter reads $\rho^{(2)}_{FF}(x_1,x_2;t)=\sum_{i=1}^{d_F^2} \lambda_i^{(2)}(t) \alpha_i^{(2)}(x_1,x_2;t)\alpha_i^{(2)*}(x_1,x_2;t)$, 
where $\lambda_i^{(2)}(t)$ and $\alpha_i^{(2)}(x_1,x_2;t)$ are the eigenvalues and eigenfunctions (also known as natural geminals) of $\rho^{(2)}_{FF}(x_1,x_2;t)$ respectively. 
Indeed, inspecting the individual contributions of $\alpha_i^{(2)}(x_1,x_2;t)$, $i=1,2,\dots$, we can deduce that the state of the 
well-separated quasiparticles is described by $\alpha_1^{(2)}(x_1,x_2;t)$ whilst the existence of the bound state is caused by the 
higher-order contributions namely $\alpha_i^{(2)}(x_1,x_2;t)$ with $i>1$ (results not shown here for brevity).

\section{Summary and Conclusions}\label{sec:conclusions}

We have investigated the nonequilibrium quantum dynamics of a harmonically trapped particle imbalanced BF mixture consisting of 
two spin-polarized fermionic impurities and a majority bosonic species upon quenching the interspecies repulsion from weak to larger values. 
Comparing the dynamics within the MF approximation and the MB level enables us to expose the crucial role of correlations, especially in 
the time-evolution of the impurities, and reveal a variety of dynamical regimes occurring for different interspecies interaction strengths. 

In the ground state of the particle imbalanced BF mixture a phase separation between the two species occurs, both at the MF and the MB level for 
increasing interspecies interaction strengths which overcome the bosonic intraspecies ones. 
In particular, the fermionic ground state single-particle density deforms from a Gaussian-like distribution located around the trap center and fully 
overlapping with the Thomas-Fermi profile of the bosonic species into two spatially separated density humps residing at the edges of this Thomas-Fermi profile. 
To trigger the dynamics we consider an interspecies interaction quench from weak towards larger repulsions. 
Depending on the final value of the interspecies interactions we realize four different dynamical regimes. 
For weak postquench interactions both species undergo a breathing motion while remaining miscible throughout the evolution. 
Here, the degree of correlations is negligible and therefore the MF approximation adequately captures the dynamics of the mixture. 
Increasing the postquench interaction strength the density of the impurities splits into two repelling density peaks moving towards the 
edges of the bosonic cloud which performs a breathing motion. 
In the MF approximation, these density branches reach the edges of the bosonic cloud and then they are reflected back to the trap center where they collide and subsequently 
exhibit a dispersive behavior. 
The latter suggests that the impurities are in a superposition of several lower-lying excited states of their external potential. 
At the MB level the dynamics of the mixture shows the same overall phenomenology. 
However, the dispersive character of the single-particle density of the impurities is more pronounced, indicating that they are in a superposition of 
higher excited states compared to the MF case. 

For stronger postquench interaction strengths the dynamical behavior of both the fermionic and the bosonic species alters drastically. 
Within the MF approximation we observe the spontaneous generation of two DB soliton states, with the bright solitary waves building 
upon the fermionic species and the corresponding dark states appearing in the bosonic gas. 
We have verified their existence by inspecting the phase jump of the dark solitons as well as by performing a fitting of the numerically obtained 
single-particle densities to the analytical waveforms of this type of states. 
Turning to the MB approach we identify significant deviations with respect to the MF dynamics. 
In particular, after the quench the fermionic density breaks into two distinct localized humps which travel towards the edges of the bosonic cloud and 
are accompanied by the formation of density dips in the latter resembling this way DB states. 
For later times, the fermionic density humps reach the edges of the bosonic cloud and undergo a small amplitude oscillatory motion around the Thomas-Fermi radius 
of the bosonic species. 
This latter process signifies the dynamical decay of the DB state and thus provides an undeniable effect of the presence of correlations. 

Quenching to very strong interspecies interactions and focusing on the MF approximation we again observe the formation of DB states 
with a slightly modified oscillation period and amplitude of the bright states when compared to weaker interspecies interactions. 
However, at the MB level the fermionic density initially splits into two repelling humps which subsequently oscillate with a very small amplitude 
and deeper in the evolution tend to approach an almost steady state. 
Here, the majority species forms density dips at the spatial regions where the density peaks of the impurities density are located. 

To further understand the dynamics we have employed other diagnostics such as the different energy contributions of each of the species as well as their 
mutual interaction energy and the one-body coherence function. 
Inspecting the energy contributions during the dynamics it becomes evident that entering the strong interspecies interaction regime after the quench an energy 
exchange process from the impurities to the bosonic bath occurs when correlations are taken into account. 
However, within the MF approximation a periodic back and forth energy transfer from the bright states (impurities) to the BEC medium takes place. 
Monitoring the one-body coherence function reveals the appearance of Mott-like correlations between the emergent fermionic density humps for strong postquench 
interactions, thus indicating their tendency for localization. 
Moreover, examining the time-evolution of the fermionic two-body reduced density matrix unveils that depending on the strength of the postquench 
repulsion the two fermions either behave almost independently or experience a weak attraction for very strong interspecies couplings. 
Finally, resorting to the population eigenvalues of the single-particle functions appearing in the MB ansatz we have identified that both inter- and intraspecies 
correlations become stronger for a larger postquench interaction strength. 

There is a multitude of several promising extensions of the current work that are of interest for future investigations. 
An interesting prospect is to unravel the corresponding interspecies interaction quench dynamics in the case of two interacting bosonic 
impurities immersed in either a bosonic or fermionic bath in order to systematically examine the role of their induced interactions in the 
course of the evolution. 
Also, the inclusion of temperature effects in such an investigation would be highly desirable \cite{Tajima,Tajima1}. 
Another intriguing direction would be to examine the periodically driven dynamics of one and two impurities inside a non-perturbed bosonic 
bath and subsequently inspect their emergent dissipative motion into the bath with respect to the driving frequency. 
Certainly the simulation of the corresponding radiofrequency spectrum \cite{Mistakidis_fermi_pol} or the structure factor of the present system \cite{Mistakidis_orthog_catastr} by considering 
spinor impurities in order to identify the emerging polaronic states and subsequently investigate their lifetime and residue constitutes an intriguing 
perspective.

\appendix

\section{Convergence and ingredients of the many-body simulations} \label{sec:convergence}

As we have already discussed in the main text, in order to simulate the correlated nonequilibrium dynamics of the considered Bose-Fermi mixture we 
utilize the Multi-Layer Multi-Configurational Time-Dependent Hartree Method for Atomic Mixtures (ML-MCTDHX) \cite{MLX}. 
It is a variational method for solving the time-dependent MB Schr{\"o}dinger equation of various types of atomic mixtures consisting either 
of bosonic \cite{mistakidis_phase_sep,Mistakidis_bose_pol} or fermionic 
\cite{Erdmann_tunel,Erdmann_phase_sep,Koutentakis_prob_fer} species and also spin degrees of freedom \cite{Mistakidis_fermi_pol,Mistakidis_orthog_catastr}. 
The key ingredient of this numerical approach is the expansion of the MB wavefunction with respect to a time-dependent 
variationally optimized basis. 
This allows us to include all the relevant intra- and interspecies correlations into our MB ansatz  
using a computationally feasible basis size. 
Therefore the relevant subspace of the Hilbert space at each time instant of the evolution is chosen in a more efficient manner when compared to methods 
relying on a time-independent basis.  

\begin{figure}
 	\includegraphics[width=0.46\textwidth]{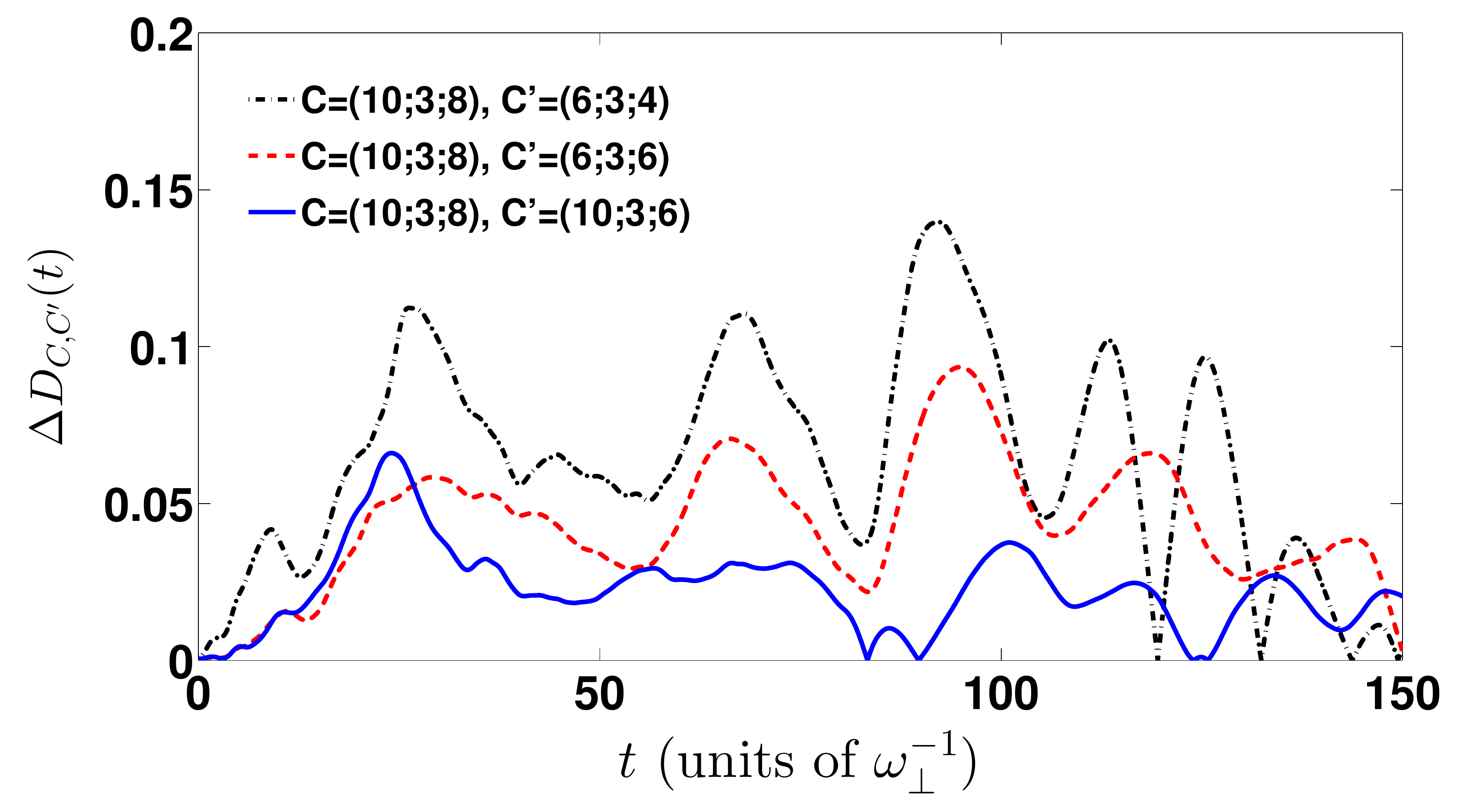}
 	\caption{Time-evolution of the deviation $\Delta D_{C,C'}(t)$ of the relative distance among the two fermionic impurities between 
 	the $C=(10;3;8)$ and other orbital configurations $C=(D;d_A;d_B)$ (see legend). 
 	The harmonically trapped BF mixture consists of $N_B=100$ bosons and $N_F=2$ fermions and it is prepared in its ground state with $g_{BB}=0.5$ and $g_{BF}=0.1$. 
 	To trigger the dynamics we follow an intersepcies interaction quench to $g_{BF}=1.5$.}
 	\label{fig:convergence} 
 \end{figure}

In particular, the considered Hilbert space truncation is designated by the employed orbital configuration space 
denoted as $C=(D;d_B;d_F)$. 
Here, $D=D_B=D_F$ and $d_B$, $d_F$ refer to the number of species and single-particle functions respectively of 
each species [see also Eqs. (\ref{Eq:WF}) and (\ref{Eq:SPF})]. 
Also, within our numerical calculations we use a primitive basis based on a sine discrete variable representation 
including 600 grid points. 
We note that this sine discrete variable representation intrinsically introduces hard-wall boundary conditions which in our case 
are imposed at $x_\pm=\pm50$ and, of course, do not affect our results since we do not observe appreciable densities to occur beyond $x_{\pm}=\pm25$. 
To infer about the convergence of the MB simulations we testify that the observables of interest become almost insensitive 
(within a given level of accuracy) upon varying the used orbital configuration space, $C=(D;d_A;d_B)$. 
Note also that all MB calculations presented in the main text are based on the orbital configuration $C=(10;3;8)$. 
To showcase the convergence of our results for a varying number of species and single-particle functions e.g. we track the relative distance 
between the two fermionic impurities during the nonequilibrium dynamics.  
More precisely, we calculate its absolute deviation between the $C=(10;3;8)$ and other orbitals configurations $C'=(D;d_A;d_B)$ 
\begin{equation}
\Delta D_{C,C'}(t) =\frac{\abs{D_C(t) -D_{C'}(t)}}{D_C(t)}. \label{deviation_rel_dist} 
\end{equation} 
Figure \ref{fig:convergence} shows $\Delta D_{C,C'}(t)$ between the two fermionic impurities upon considering an interspecies interaction quench 
from $g_{BF}=0.1$ to $g_{BF}=1.5$.    
It becomes evident that a systematic convergence of $\Delta D_{C,C'}(t)$ is achieved. 
For instance, comparing $\Delta D_{C,C'}(t)$ between the $C=(10;3;8)$ and $C'=(6;3;6)$ orbital configurations we deduce that the corresponding 
relative difference lies below $9\%$ throughout the evolution. 
Also, the relative deviation, $\Delta D_{C,C'}(t)$, when $C=(10;3;8)$ and $C'=(10;3;6)$ becomes at most of the order of $6\%$ in the course of the dynamics. 
Finally, we remark that a similar analysis has been performed for all other postquench interspecies interaction strengths presented in the main text 
and found to be adequately converged (not shown here for brevity).

\section*{Acknowledgements} 
S.I.M. and P.S. gratefully acknowledge financial support by the Deutsche Forschungsgemeinschaft 
(DFG) in the framework of the SFB 925 ``Light induced dynamics and control of correlated quantum
systems''. 
S.I.M thanks A. G. Volosniev for illuminating discussions regarding the polaron problem.

{}

\end{document}